\RequirePackage{tikz-cd}
\documentclass[sn-mathphys-num]{sn-jnl}
\usepackage{lineno,hyperref}
\usepackage{url}
\usepackage{stmaryrd}
\usepackage{tikz-cd}
\usepackage{yfonts}
\usepackage[utf8]{inputenc}
\usepackage[T1]{fontenc}
\usepackage [autostyle]{csquotes}
\usepackage{caption}
\usepackage{amsfonts}
\usepackage{amsthm}
\usepackage{eqnarray}
\usepackage{mathtools}

\usepackage{float}
\usepackage{amssymb}
\usepackage{amsmath}
\usepackage{enumerate}
\usepackage{lipsum}
\theoremstyle{thmstyleone}
\newtheorem{thm}{Theorem}%[section]
\newtheorem{lem}{Lemma}
\newtheorem{prop}{Proposition}
\newtheorem{cor}{Corollary}
\newtheorem{defn}{Definition}%[section]
\newtheorem{rem}{Remark}
\newtheorem{note}{Note}
\newtheorem{fact}{Fact}
\begin{document}
%	\begin{frontmatter}
	\title{Duality for Fitting’s Multi-valued Modal logic via bitopology and bi-Vietoris coalgebra}{}
	\author*[1,2]{\fnm{Litan Kumar} \sur{Das}}\email{ld06iitkgp@gmail.com}
	\author[2,3]{\fnm{Kumar Sankar} \sur{Ray}}\email{ksray@isical.ac.in}
	\equalcont{These authors contributed equally to this work.}
	\author[1,2]{\fnm{Prakash Chandra} \sur{Mali}}\email{pcmali1959@gmail.com}
	\equalcont{These authors contributed equally to this work.}
	\affil*[1]{\orgdiv{Department of Mathematics}, \orgname{Jadavpur University}, \orgaddress{\street{Jadavpur}, \city{Kolkata}, \postcode{700032}, \state{West Bengal}, \country{India}}}
	
	\affil[2]{\orgdiv{ECSU}, \orgname{Indian Statistical Institute, Kolkata}, \orgaddress{\street{}, \city{Kolkata}, \postcode{700108}, \state{West Bengal}, \country{India}}}
	
	\affil[3]{\orgdiv{Department of Mathematics}, \orgname{Jadavpur University}, \orgaddress{\street{Jadavpur}, \city{Kolkata}, \postcode{700032}, \state{West Bengal}, \country{India}}}
%\author[1]{Litan Kumar Das\corref{cor1}}
%\ead{ld06iitkgp@gmail.com}
%\author[2]{Kumar Sankar Ray}
%\ead{ksray@isical.ac.in}
%\author[3]{Prakash Chandra Mali}
%\ead{pcmali1959@gmail.com}
%\ead{ld06iitkgp@gmail.com}
%\affiliation[1]{organization={Department of Mathematics, Jadavpur University}, addressline={Jadavpur},
%	postcode={700032}, city={Kolkata}, country={India}}
	
%	\affiliation[2]{organization={ECSU, Indian Statistical Institute, Kolkata}, 
	%	postcode={700108},  country={India}}
		
	%	\affiliation[3]{organization={Department of Mathematics, Jadavpur University}, 
		%	addressline={Jadavpur},
		%	postcode={700032}, city={Kolkata}, country={India}}
%\address[addresstwo]{Some other Address.}
%\begin{abstract}
\abstract{Fitting’s Heyting-valued logic and Heyting-valued modal logic have already been studied from an algebraic viewpoint. In addition to algebraic axiomatizations with the completeness of Fitting’s Heyting-valued logic and Heyting-valued modal logic, both topological and coalgebraic dualities have also been developed for algebras of Fitting's Heyting-valued modal logic. Bitopological methods have recently been employed to investigate duality for Fitting's Heyting-valued logic. However, the concepts of bitopology and bi-Vietoris coalgebras are conspicuously absent from the development of dualities for Fitting's many-valued modal logic. With this study, we try to bridge that gap. The main results are bitopological and coalgebraic duality for Fitting's many-valued modal logic. We develop a bitopological duality for algebras of Fitting’s Heyting-valued modal logic by extending known bitopological duality for Fitting's non-modal logic. To develop coalgebraic duality, we adapt Lauridsen's bi-Vietoris construction from the category of pairwise Stone spaces to the category $PBS_{\mathcal{L}}$ of $\mathcal{L}$-valued (with $\mathcal{L}$ a bounded finite distributive lattice, i.e., a Heyting algebra) pairwise Boolean spaces by incorporating a structure map, and from this obtain the $\mathcal{L}$-biVietoris functor.
	 Finally, we establish dual equivalence between coalgebras for the $\mathcal{L}$-biVietoris functor and algebras of Fitting’s $\mathcal{L}$-valued modal logic. As a result, we conclude that Fitting’s Heyting-valued modal logic is sound and complete with respect to the coalgebras of $\mathcal{L}$-biVietoris functor. We also apply this coalgebraic approach to the bitopological duality to show the existence of cofree and final coalgebras, and to establish a Hennessy-Milner property. }
%	\end{abstract}
	%	\begin{keyword}
	%	\end{keyword}
%	\end{frontmatter}
	\maketitle
	\section{Introduction}
	The algebraic axiomatization of a modified version of Fitting’s Heyting-valued modal logic has already been addressed in \cite{maruyama2009algebraic}. Maruyama \cite{maruyama2011dualities} proposed J{\'o}nsson-Tarski topological duality (see \cite{blackburn2001modal,hansoul1983duality,jonsson1951boolean}) for $\mathcal{L}$-$\mathcal{ML}$-algebras (algebras of Fitting’s Heyting-valued modal logic), while \cite{maruyama2012natural} developed both topological and coalgebraic dualities within a natural duality and coalgebra framework for general algebraic structures encompassing algebras of Fitting's many-valued modal logic.\\
	
	An exemplary story in coalgebraic logic can be found in \cite{kurz2006coalgebras}. The Stone duality \cite{stone1938representation} between Boolean algebras and sets represents the syntax and semantics of a propositional logic. The algebras and coalgebras of the endofunctors define the syntax and semantics of the modal propositional logic. As an illustration, the modal logic $K$ and Kripke semantics derive from the Stone duality by taking an endofunctor on sets. So, in acceptable circumstances, we can achieve duality between the relevant algebras and coalgebras. In addition to demonstrating the fact that the widely recognised Stone duality could be articulated in coalgebraic terms, Abramsky \cite{abramsky2011cook} discovered a coalgebraic approach to the J{\'o}nsson-Tarski duality for modal algebras (for more information, see also \cite{kupke2004stone}). It is specifically noted that the category of Boolean spaces is isomorphic to the category of descriptive general Kripke frames. Esakia \cite{esakia1974topological} also noticed this connection. Coalgebras for the Vietoris functor on the category of Boolean spaces, also known as Stone spaces, can thus be used to represent semantics for modal logic. In \cite{palmigiano2004coalgebraic}, the author showed that coalgebras of a Vietoris functor on the category of Priestley spaces, i.e., compact, totally ordered disconnected spaces, provide sound and complete semantics for positive modal logic. The author in \cite{lauridsen2015bitopological} subsequently developed a Vietoris functor on the category of pairwise Stone spaces, whereby coalgebras for that defined functor yield sound and complete semantics for the positive modal logic.\\
	
	Although these existing strands independently address substantial aspects of the field, Maruyama \cite{maruyama2012natural} offers a natural duality and coalgebraic framework sufficiently broad to incorporate algebras of Fitting's many-valued modal logic, Priestley-based hyperspace semantics for positive modal logics have been integrated by Bezhanishvili et al. \cite{bezhanishvili2023remarks}, and Lauridsen \cite{lauridsen2015bitopological} presents a bi-Vietoris construction and coalgebraic completeness within the Pairwise Stone space context- a considerable gap persists. There is no existing literature that defines a duality of many-valued modal logic based on the techniques of bitopology. It is important to mention that there is a conspicuous lack of an explicit coalgebraic semantics based on a bitopological framework of many-valued modal logic. Furthermore, it does not provide a formal demonstration that this coalgebraic category is dual to the category of algebras of Fitting-style many-valued modal logic.\\
	This paper fills that gap. We integrate these methodologies to develop bitopological and coalgebraic dualities for $\mathcal{L}$-$\mathcal{ML}$-algebras, where $\mathcal{L}$ is a semi-primal algebra having a bounded lattice reduct. Our general theory extends both the J{\'o}nsson-Tarski duality and the Abramsky-Kupke-Kurz-Venema coalgebraic duality \cite{abramsky2011cook,kupke2004stone} in the setting of bitopological language. \\
	We aim to develop a bitopological duality for algebras of Fitting’s Heyting-valued modal logic by setting up a notion of $PRBS_{\mathcal{L}}$ as a category of $\mathcal{L}$-valued pairwise Boolean spaces equipped with a structure map and a  compatible binary relation. As a result, the natural duality theory and J{\'o}nsson-Tarski duality for modal algebras are expanded within the framework of bitopological languages, thereby generalizing the bitopological duality for Fitting's $\mathcal{L}$-valued logic \cite{das2021bitopological} to the modal case. \\
	Building upon the bitopological duality for Fitting's $\mathcal{L}$-valued modal logic, we derive a coalgebraic semantics for Fitting's many-valued modal logic. The adaptation of Lauridsen's \cite{lauridsen2015bitopological} bi-Vietoris functor on the category of pairwise Stone spaces serves as the cornerstone of this development. We then aim to make this machinery fit the many-valued context by defining an endofunctor, denoted by $V_{\mathcal{L}}^{bi}$ on the category $PBS_{\mathcal{L}}$ of $\mathcal{L}$-valued pairwise Boolean spaces endowed with a structure map. This ensures the functor $V_{\mathcal{L}}^{bi}$ preserves the $\mathcal{L}$-valued structure that is necessary in handling the many-valued semantics. We do not claim that the bi-Vietoris mechanism is novel; rather, we adapt it in a many-valued environment and change the algebraic semantics to coalgebraic semantics in bitopological setting by establishing the coalgebraic duality theory.\\
	This study provides the first coherent bitopological and coalgebraic structure of Fitting's many-valued modal logic. Our main contributions are:
	\begin{itemize}
		\item A bitopological duality between the category of algebras of Fitting's many-valued modal logic and a category $PRBS_{\mathcal{L}}$ of $\mathcal{L}$ valued relational pairwise Boolean spaces.
		\item A coalgebraic semantics based on a specified $\mathcal{L}$-biVietoris endofunctor on the category $PBS_{\mathcal{L}}$, revealing that the category of coalgebras of this  $\mathcal{L}$-biVietoris endofunctor is isomorphic to $PRBS_{\mathcal{L}}$. This combination of algebra, bitopology, and coalgebra provides us with new, valuable insights for Fitting's many-valued modal logic, including soundness, completeness, the Hennessy-Milner property for image-finite models, and the existence of final and cofree coalgebras.
	\end{itemize}
	\textbf{Significance and connection to the existing dualities:}\\
	The framework of Maruyama's study  \cite{maruyama2012natural} provides an extensive natural duality and coalgebraic structure that encompasses the algebras of Fitting's many-valued modal logic. Our findings agree with that paradigm but with a bitopological improvement that is specifically tailored to the multi-valued setting.
	%while introducing a bitopological enhancement specifically designed for the multi-valued context.
	 In the case that we ignore the second topology and change the structure map in the standard topological setting, our framework is then the one discussed by Maruyama \cite{maruyama2012natural}. Consequently, the dualities proposed in our work constitute a conservative and structured enhancement of that general setting.\\
	Maruyama's work \cite{maruyama2011dualities} establishes the J{\'o}nsson-Tarski duality for algebras of Fitting’s $\mathcal{L}$-valued modal logic within a standard topological framework. Our study adapts those methods to models characterised as $\mathcal{L}$-valued relational pairwise Boolean spaces, i.e., pairwise Boolean spaces equipped with a subalgebra-indexed meet-preserving structure map and a binary relation that is compatible with both the topologies and the structure map, thus developing a bitopological duality for the modal case. This methodology does not employ Vietoris machinery; rather, it is a direct adaptation of J{\'o}nsson-Tarski and natural duality to bitopological languages.\\
	In \cite{lauridsen2015bitopological}, Lauridsen developed the bi-Vietoris construction and coalgebraic completeness for positive modal logic within the framework of pairwise Stone spaces. Our study (see Section \ref{COALGD}) lifts that bi-Vietoris construction to the many-valued context: we define a $\mathcal{L}$-biVietoris functor on the category $PBS_{\mathcal{L}}$ of $\mathcal{L}$-valued pairwise Boolean spaces endowed with a structure map. The $\mathcal{L}$-biVietoris functor preserves the $\mathcal{L}$-valued structure, and its coalgebras coincide with our $\mathcal{L}$-valued relational model. Thus, our adaptation is non-trivial.\\
	If we consider $\mathcal{L}=\mathbf{2}$, the two-element Boolean algebra, our framework reduces to the classical case: the structure map becomes trivial and on the canonical dual spaces, $\mathfrak{G}(\mathcal{A})=(HOM_{\mathcal{VA}_{\mathcal{L}}}(\mathcal{A},\mathcal{L}),\tau_1,\tau_2,\alpha_{\mathcal{A}})$ that are obtained by the duality functor (cf. Section \ref{BDFML}), both the topologies $\tau_1$ and $\tau_2$ have the same basis, i.e., $\tau_1=\tau_2$.
	As a result, the category $PBS_{\mathbf{2}}$ matches the category of classical Stone spaces. Thus, in the two-valued case, our $\mathcal{L}$-biVietoris functor reflects the Lauridsen's bi-Vietoris functor on pairwise Stone spaces with equal topologies. \\
	\textbf{Assumptions and Scope}\\
	Through out this study, we assume $\mathcal{L}$ as a truth-value set which is bounded finite distributive lattice, i.e., a Heyting algebra. We work with pairwise Boolean spaces equipped with a subalgebra-indexed meet preserving structure map and algebras of Fitting's $\mathcal{L}$-valued modal logic. \\
	Assuming this, we modify bi-Vietoris construction of Lauridsen to define the concept of $\mathcal{L}$-biVietoris functor $V_{\mathcal{L}}^{bi}$, work out both bitopological and coalgebraic dulity theorems for $\mathcal{L}$-valued modal logic, and generate the fundamental consequences: soundness and completeness, existence of final, cofree coalgebras, and Hennessy-Milner property on image-finite coalgebras of our adapted bi-Vietoris functor $V_{\mathcal{L}}^{bi}$.

The structure of the paper is as follows:\\
 In Section \ref{PRE}, we review the fundamental notions of bitopological spaces and algebras of Fitting’s Heyting-valued logic, denoted as $\mathcal{L}$-$\mathcal{VL}$-algebras, and we also discuss the bitopological duality theorem for $\mathcal{L}$-$\mathcal{VL}$-algebras. Section \ref{BDFML} contains our first main result: a bitopological duality for Fitting’s Heyting-valued modal logic.  Section \ref{COALGD} extends Lauridsen's bi-Vietoris construction from the category of pairwise Stone spaces to the category of $\mathcal{L}$-valued pairwise Boolean spaces, $PBS_{\mathcal{L}}$, resulting in an endofunctor $V_{\mathcal{L}}^{bi}$ on the category $PBS_{\mathcal{L}}$ that preserves the many-valued structure. We identify the category of $\mathcal{L}$-valued relational pairwise Boolean spaces ,$PRBS_{\mathcal{L}}$, with the category $COALG(V_{\mathcal{L}}^{bi})$ of coalgebras of the functor $V_{\mathcal{L}}^{bi}$, and by combining this with the result in Section \ref{BDFML}, we obtain the coalgebraic duality for Fitting’s Heyting-valued modal logic and this yields soundness and completeness of Fitting's many-valued modal logic with respect to $V_{\mathcal{L}}^{bi}$-coalgebras.  In Section \ref{LAB}, we provide the applications of our coalgebraic approach to the bitopological duality. The fundamental applications are:
 \begin{itemize}
 \item A demonstration of the Hennessy-Milnar property for $V_{\mathcal{L}}^{bi}$-coalgebras.
 \item The existence of cofree coalgebras, and final coalgebras.
\end{itemize}
  In Section \ref{FCON}, we conclude the paper by indicating the limitations of our study and 
  provide some possible avenues of future research.
\section{Preliminaries}
\label{PRE}	
	The readers are believed to be conversant with the fundamental ideas of category theory and topology. For details on universal algebra and lattice theory, we refer the reader to \cite{burris1981course,davey2002introduction}. To get more information about category theory, see \cite{adamek1990abstract}.
\subsection{Bitopological spaces}
A triple $(X,\tau_1,\tau_2)$, where $(X,\tau_1)$ and $(X,\tau_2)$ are topological spaces, is called a bitopological space. Consider $\delta_1$ and $\delta_2$ represent, respectively, the collections of $\tau_1$-closed sets and $\tau_2$-closed sets. We set $\beta_1=\tau_1\cap\delta_2$ and $\beta_2=\tau_2\cap\delta_1$.
\begin{defn}[\cite{salbany1974bitopological}]
\label{BT}
\begin{enumerate}[(i)]
\item A bitopoological space $(X,\tau_1,\tau_2)$ is said to be pairwise Hausdorff space if for every pair $(x,y)$ of distinct points $x,y\in X$ there exists disjoint open sets $U_x\in\tau_1$ and $U_y\in\tau_2$ containing $x$ and $y$, respectively.
\item A bitopoological space $(X,\tau_1,\tau_2)$ is said to be pairwise zero-dimensional if $\beta_1$ is a basis for $\tau_1$ and $\beta_2$ is a basis for $\tau_2$.
\item A bitopoological space $(X,\tau_1,\tau_2)$ is said to be pairwise compact if the topological space $(X,\tau)$, where $\tau=\tau_1\vee\tau_2$, is compact.
\end{enumerate}
\end{defn}
According to Alexander’s Lemma( a classical result in general topology), the idea of pairwise compactness described in Definition \ref{BT} is equivalent to the condition that every cover $\{U:U\in\tau_1\cup\tau_2\}$ of X has a finite subcover. A pairwise Boolean space is a bitopological space that is pairwise Hausdorff, pairwise zero-dimensional, and pairwise compact. A map $f:(P,\tau_1,\tau_2)\to (P_1,\tau_1^1,\tau_2^1)$ is said to be pairwise continuous if the map $f:(P,\tau_i)\to (P_1,\tau_i^1)$ is continuous for $i\in\{1,2\}$. Pairwise Boolean spaces and pairwise continuous maps form a category, denoted by $PBS$.
\begin{prop}[\cite{lauridsen2015bitopological}]
\label{PROP1}
If $T_1$ and $T_2$ are subbasis for the topologies $\tau_1$ and $\tau_2$, respectively, then $T_1\cup T_2$ is a subbasis for the topology $\tau_1\vee\tau_2$.
\end{prop}
\begin{prop}[\cite{lauridsen2015bitopological}]
\label{PROP2}
Let $(X,\tau_1,\tau_2)$ be a pairwise compact bitopological space. 
Consider a finite collection $\{C_i:C_i\in\delta_1\cup\delta_2, i=1,2,\cdots,n\}$ of subsets of $X$. Then $\displaystyle\bigcap_{i=1}^nC_i$ is pairwise compact.
\end{prop}
It is clear from the above proposition that any $\tau_1$-closed or $\tau_2$-closed subset of a pairwise compact space $X$ is pairwise compact.
\subsection{Fitting’s Heyting-valued logics}
Fitting \cite{fitting1991many} proposed $\mathcal{L}$-valued logics and $\mathcal{L}$-valued modal logics for a finite distributive lattice $\mathcal{L}$(i.e., $\mathcal{L}$ is a Heyting algebra) in 1991. Maruyama \cite{maruyama2009algebraic} introduced algebraic axiomatization of Fitting’s logics. In \cite{maruyama2009algebraic} the author studied Fitting’s Heyting-valued logic and Heyting-valued modal logic without regard for fuzzy truth constants other than $0$ and $1$, and added a new operation $T_{\ell}(-)$, $\ell\in\mathcal{L}$. From a logical perspective, $T_{\ell}(p)$ infers that the truth value of a proposition $p$ is $\ell$. The operations of $\mathcal{L}$-valued logic, denoted by $\mathcal{L}$-$\mathcal{VL}$, are $\vee,\wedge,\rightarrow,0,1$ and $T_{\ell}(-)$, $\ell\in\mathcal{L}$, where $\vee,\wedge,\rightarrow$ are binary operations, $0$ and $1$ are nullary operations and $T_{\ell}$ is a unary operation. For $\ell_1,\ell_2\in\mathcal{L}$, $\ell_1\rightarrow\ell_2$ means the pseudo-complement of $\ell_1$ relative to $\ell_2$.\\
The following lemmas describe some term functions.
\begin{lem}\label{TF1}
Define a function $f:\mathcal{L}^4\to\mathcal{L}$ by 
\[ f(\ell_1,\ell_2,\ell_3,\ell_4)=  \left\{
\begin{array}{ll}
\ell_3 & (\ell_1=\ell_2)\\
\ell_4 & (\ell_1\neq\ell_2) \\
\end{array} 
\right. \]
Then, $f$ is a term function of $\mathcal{L}$.
\end{lem}
\begin{lem}\label{TF2}
For every $\ell\in\mathcal{L}$, define $T_{\ell}:\mathcal{L}\to\mathcal{L}$ by 
\[ T_{\ell}(\ell')=  \left\{
\begin{array}{ll}
1 & (\ell'=\ell)\\
0 & (\ell'\neq\ell) \\
\end{array} 
\right. \]
Then, $T_{\ell}$ is a term function of $\mathcal{L}$.
\end{lem}
\begin{lem} 
\label{TF3}
Let $\ell\in\mathcal{L}$. Then the function $U_{\ell}:\mathcal{L}\to\mathcal{L}$ defined by 
\[ U_{\ell}(\ell')=  \left\{
\begin{array}{ll}
1 & (\ell'\geq\ell)\\
0 & (\ell'\ngeq\ell) \\
\end{array} 
\right. \]
, is a term function of $\mathcal{L}$.
\end{lem}
We recall the idea of $\mathcal{L}$-$\mathcal{VL}$-algebras, which provides sound and complete semantics of $\mathcal{L}$-valued logic $\mathcal{L}$-$\mathcal{VL}$.
\begin{defn}[\cite{maruyama2009algebraic}]
\label{LVL}
An algebraic structure $(\mathcal{A},\wedge,\vee,\rightarrow,T_{\ell}(\ell\in\mathcal{L}),0,1)$ is said to be a $\mathcal{L}$-$\mathcal{VL}$-algebra iff the following conditions hold for any  $\ell_1, \ell_2\in\mathcal{L}$, and $a,b\in\mathcal{A}$:
\begin{enumerate}[(i)]
\item $(\mathcal{A},\wedge,\vee,\rightarrow,T_{\ell}(\ell\in\mathcal{L}),0,1)$ is a Heyting algebra;
\item $T_{\ell_1}(a)\wedge T_{\ell_2}(b)\leq T_{\ell_1\rightarrow \ell_2}(a\rightarrow b)\wedge T_{\ell_1\wedge \ell_2}(a\wedge b)\wedge T_{\ell_1\vee \ell_2}(a\vee b)$;\\
$T_{\ell_2}(a)\leq T_{T_{\ell_1}(\ell_2)}(T_{\ell_1}(a))$;
\item $T_0(0)=1$; $T_{\ell}(0)=0$ ($\ell\neq 0$); $T_1(1)=1$; $T_{\ell}(1)=0$, if $\ell\neq 1$;
\item $\bigvee\{T_{\ell}(a): \ell\in\mathcal{L}\}=1$; $T_{\ell_1}(a)\vee (T_{\ell_2}(a)\rightarrow 0)=1$;\\
$T_{\ell_1}(a)\wedge T_{\ell_2}(a)=0$, $(\ell_1\neq \ell_2)$;
\item $T_1(T_{\ell}(a))=T_{\ell}(a)$, $T_0(T_{\ell}(a))=T_{\ell}(a)\rightarrow 0$, $T_{\ell_2}(T_{\ell_1}(a))=0$, $(\ell_2\neq 0,1)$;
\item $T_1(a)\leq a$, $T_1(a\wedge b)=T_1(a)\wedge T_1(b)$;
\item $\displaystyle\bigwedge_{\ell\in\mathcal{L}}(T_{\ell}(a)\leftrightarrow T_{\ell}(b))\leq (a\leftrightarrow b)$.
\end{enumerate}
\end{defn}
\begin{note}
The class of all $\mathcal{L}$-$\mathcal{VL}$-algebras forms a variety (in the sense of universal algebra). If $\mathcal{L} = \{0, 1\}$, then $\mathcal{L}$-$\mathcal{VL}$-algebras becomes Boolean algebras.
\end{note}
\begin{defn}\label{LVLHOM}
A function between $\mathcal{L}$-$\mathcal{VL}$-algebras is said to be homomorphism if it preserves the operations $\vee,\wedge,\rightarrow,T_{\ell}(\ell\in\mathcal{L}),0,1$.
\end{defn}
Let $\mathcal{VA}_{\mathcal{L}}$ denote the category of $\mathcal{L}$-$\mathcal{VL}$-algebras. \\
$\mathcal{L}$-valued modal logic denoted by $\mathcal{L}$-$\mathcal{ML}$, is defined by $\mathcal{L}$-valued Kripke semantics. The idea of $\mathcal{L}$-valued Kripke semantics can be found in \cite{maruyama2011dualities}. The operations of $\mathcal{L}$-valued modal logic $\mathcal{L}$-$\mathcal{ML}$ are the operations of $\mathcal{L}$-$\mathcal{VL}$ and a unary operation $\Box$, called modal operation. We now recall the concept of $\mathcal{L}$-$\mathcal{ML}$-algebras, which define a sound and complete semantics for $\mathcal{L}$-$\mathcal{ML}$.
\begin{defn}[\cite{maruyama2009algebraic}]
\label{LML}
An algebraic structure $(\mathcal{A},\wedge,\vee,\rightarrow,T_{\ell}(\ell\in\mathcal{L}),\Box,0,1)$ is said to be a $\mathcal{L}$-$\mathcal{ML}$-algebra iff it satisfies the following conditions:
\begin{enumerate}[(i)]
\item $(\mathcal{A},\wedge,\vee,\rightarrow,T_{\ell}(\ell\in\mathcal{L}),0,1)$ is a $\mathcal{L}$-$\mathcal{VL}$-algebra;
\item $\Box(a\wedge b)=\Box a\wedge\Box b$;
\item $\Box U_{\ell}(a)=U_{\ell}(\Box a)$, $\forall \ell\in\mathcal{L}$, where the unary operation $U_{\ell}(\ell\in\mathcal{L})$ is defined by $U_{\ell}(a)=\bigvee\{T_{\ell'}(a):\ell\leq\ell',\ell'\in\mathcal{L}\}$, $a\in\mathcal{A}$. Logically, it means that the truth value of $a$ is greater than or equal to $\ell$.
\end{enumerate}
\end{defn}
A homomorphism of $\mathcal{L}$-$\mathcal{ML}$-algebras is a function that preserves all the operations of $\mathcal{L}$-$\mathcal{VL}$-algebras and the modal operation $\Box$. Let $\mathcal{MA}_{\mathcal{L}}$ denote the
category of $\mathcal{L}$-$\mathcal{ML}$-algebras and homomorphisms of $\mathcal{L}$-$\mathcal{ML}$-algebras.\\
For a Kripke frame $(P,\mathcal{R})$, $\mathcal{R}[x]=\{y\in P:x\mathcal{R}y\}$, where $x\in P$, and $\mathcal{R}^{-1}[P']=\{y\in P:\exists x\in P', y\mathcal{R}x\}$, where $P'\subseteq P$. We recall a modal operation $\Box_{\mathcal{R}}$ on $\mathcal{L}$-valued powerset algebra $\mathcal{L}^P$ of $P$.
\begin{defn}[\cite{maruyama2009algebraic}]
\label{MO}
Let $(P,\mathcal{R})$ be a Kripke frame and $f\in\mathcal{L}^P$. Then $\Box_{\mathcal{R}}f:P\to\mathcal{L}$ is defined by $(\Box_{\mathcal{R}}f)(x)=\bigwedge\{f(y):y\in\mathcal{R}[x]\}$.
\end{defn}
\begin{defn}[\cite{maruyama2011dualities}]
\label{BR}
Let $\mathcal{A}$ be an object in $\mathcal{MA}_{\mathcal{L}}$. A binary relation $\mathcal{R}_{\Box}$ on $HOM_{\mathcal{VA}_{\mathcal{L}}}(\mathcal{A},\mathcal{L})$ is defined as follows:\\
$\psi\mathcal{R}_{\Box}\phi\iff \forall\ell\in\mathcal{L},\forall a\in\mathcal{A}, \psi(\Box a)\geq\ell\Rightarrow \phi(a)\geq\ell$.
\end{defn}
A $\mathcal{L}$-valued map $\mathcal{D}:HOM_{\mathcal{VA}_{\mathcal{L}}}(\mathcal{A},\mathcal{L})\times\mathcal{A}\to\mathcal{L}$ is defined by $\mathcal{D}(\psi,a)=\psi(a), \psi\in HOM_{\mathcal{VA}_{\mathcal{L}}}(\mathcal{A},\mathcal{L})$.
\begin{lem}[\cite{maruyama2011dualities}]
\label{KM}
The $\mathcal{L}$-valued canonical model $(HOM_{\mathcal{VA}_{\mathcal{L}}}(\mathcal{A},\mathcal{L}),\mathcal{R}_{\Box},\mathcal{D})$ of $\mathcal{A}$ is a $\mathcal{L}$-valued Kripke model. Then, $\mathcal{D}(\psi,\Box a)=\psi(\Box a)=\bigwedge\{\phi(a):\phi\in\mathcal{R}_{\Box}[\psi]\}$.
\end{lem}
\subsection{Bitopological duality for Fitting’s Heyting-valued logic}
We will introduce the key ideas and findings from the bitopological duality
theory for Fitting’s Heyting-valued logic. We refer to \cite{das2021bitopological} for a more thorough explanation of the bitopological duality for Fitting’s Heyting-valued logic. Let $\mathfrak{S}_{\mathcal{L}}$ denote the collection of subalgebras of $\mathcal{L}$. For a pairwise Boolean space $\mathcal{B}$, $\Lambda_{\mathcal{B}}$ denotes the collection of pairwise closed subspaces of $\mathcal{B}$. It is shown in \cite{adnadzhevich1987bicompactness} that a pairwise closed subset of a pairwise compact space is also pairwise compact. Hence, each member of $\Lambda_{\mathcal{B}}$ is a pairwise Boolean space. A finite distributive lattice $\mathcal{L}$ endowed with unary operation $T_{\ell}(\ell\in\mathcal{L})$ forms a semi-primal algebra. We have expanded the theory of natural duality \cite{clark1998natural} by creating a bitopological duality for $\mathcal{L}$-$\mathcal{VL}$-algebras \cite{das2021bitopological}.\\
We now recall the category $PBS_{\mathcal{L}}$ from \cite{das2021bitopological}.
\begin{defn}[\cite{das2021bitopological}]
\label{PBSL}
The category $PBS_{\mathcal{L}}$ is defined as follows:
\begin{enumerate}[(1)]
\item Objects$\colon$ An object in $PBS_{\mathcal{L}}$ is a tuple $(\mathcal{B},\alpha_{\mathcal{B}})$ where $\mathcal{B}$ is a pairwise Boolean space and a mapping $\alpha_{\mathcal{B}}:\mathfrak{S}_{\mathcal{L}}\to\Lambda_{\mathcal{B}}$ satisfies the following conditions:
\begin{enumerate}[(i)]
\item $\alpha_{\mathcal{B}}(\mathcal{L})=\mathcal{B}$;
\item if $\mathcal{L}_1=\mathcal{L}_2\wedge\mathcal{L}_3 (\mathcal{L}_1,\mathcal{L}_2,\mathcal{L}_3\in\mathcal{L})$, then $\alpha_{\mathcal{B}}(\mathcal{L}_1)=\alpha_{\mathcal{B}}(\mathcal{L}_2)\cap\alpha_{\mathcal{B}}(\mathcal{L}_3)$.
\end{enumerate}
\item Arrows$\colon$ An arrow $\psi:(\mathcal{B}_1,\alpha_{\mathcal{B}_1})\to(\mathcal{B}_2,\alpha_{\mathcal{B}_2})$ in $PBS_{\mathcal{L}}$ is a pairwise continuous map $\psi:\mathcal{B}_1\to\mathcal{B}_2$ that satisfies the criterion that if $x\in\alpha_{\mathcal{B}_1}(\mathcal{L}_1)(\mathcal{L}_1\in\mathfrak{S}_{\mathcal{L}})$, then $\psi(x)\in\alpha_{\mathcal{B}_2}(\mathcal{L}_1)$ i.e., $\psi$ is a subspace preserving map.
\end{enumerate}
\end{defn}
\begin{note}
\begin{enumerate}[(1)]
\item The bitopological space $(\mathcal{L},\tau,\tau)$, where $\tau$ is the discrete topology on $\mathcal{L}$, is a pairwise Boolean space. Hence, $(\mathcal{L},\tau,\tau,\alpha_{\mathcal{L}})$, where $\alpha_{\mathcal{L}}$ is a mapping from $\mathfrak{S}_{\mathcal{L}}$ to $\Lambda_{\mathcal{L}}$ that is defined by $\alpha_{\mathcal{L}}(\mathcal{L}')=\mathcal{L}'$, is an object in $PBS_{\mathcal{L}}$.
\item For an object $\mathcal{A}$ in $\mathcal{VA}_{\mathcal{L}}$ , consider a bitopological space $(HOM_{\mathcal{VA}_{\mathcal{L}}}(\mathcal{A},\mathcal{L}),\tau_1,\tau_2)$, where the topologies $\tau_1$ and $\tau_2$ are generated by the bases $B^{\tau_1}=\{\langle a\rangle:a\in\mathcal{A}\}$, where $\langle a\rangle=\{h\in HOM_{\mathcal{VA}_{\mathcal{L}}}(\mathcal{A},\mathcal{L}):h(a)=1\}$, and $B^{\tau_2}=\{B^c:B\in B^{\tau_1}\}$, respectively. Here, $B^c$ denotes the complement of $B$.
\end{enumerate}
\end{note}
\begin{fact}[\cite{das2021bitopological}]
The bitopological space $(HOM_{\mathcal{VA}_{\mathcal{L}}}(\mathcal{A},\mathcal{L}),\tau_1,\tau_2)$ is a pairwise Boolean space.
\end{fact}
The duality between the categories $\mathcal{VA}_{\mathcal{L}}$ and $PBS_{\mathcal{L}}$ is obtained via the following functors.
\begin{defn}[\cite{das2021bitopological}]
\label{FUNC1}
A contravariant functor $\mathfrak{F}:PBS_{\mathcal{L}}\to\mathcal{VA}_{\mathcal{L}}$ is defned as follows:
\begin{enumerate}[(i)]
\item For an object $(\mathcal{B},\alpha_{\mathcal{B}})$ in $PBS_{\mathcal{L}}$, define $\mathfrak{F}(\mathcal{B},\alpha_{\mathcal{B}})=(HOM_{PBS_{\mathcal{L}}}((\mathcal{B},\alpha_{\mathcal{B}}),(\mathcal{L},\alpha_{\mathcal{L}})),\vee,\wedge,\rightarrow,T_{\ell}(\ell\in\mathcal{L}),0,1)$, where $\vee,\wedge,\rightarrow,T_{\ell}(\ell\in\mathcal{L}),0,1$ are pointwise operations on the set $HOM_{PBS_{\mathcal{L}}}((\mathcal{B},\alpha_{\mathcal{B}}),(\mathcal{L},\alpha_{\mathcal{L}}))$. The operations $0$ and $1$ are regarded as constant functions, with $0$ and $1$ being their respective values.
\item For an arrow $\phi:(\mathcal{B},\alpha_{\mathcal{B}})\to (\mathcal{B}',\alpha_{\mathcal{B}'})$ in $PBS_{\mathcal{L}}$, define $\mathfrak{F}(\phi):\mathfrak{F}((\mathcal{B}',\alpha_{\mathcal{B}'}))\to \mathfrak{F}((\mathcal{B},\alpha_{\mathcal{B}}))$ by $\mathfrak{F}(\phi)(\zeta)=\zeta\circ\phi$, where $\zeta\in HOM_{PBS_{\mathcal{L}}}((\mathcal{B}',\alpha_{\mathcal{B}'}),(\mathcal{L},\alpha_{\mathcal{L}}))$.
\end{enumerate}
\end{defn}
\begin{defn}[\cite{das2021bitopological}]
\label{FUNC2}
A contravariant functor $\mathfrak{G}:\mathcal{VA}_{\mathcal{L}}\to PBS_{\mathcal{L}}$ is defined as follows:
\begin{enumerate}[(i)]
\item $\mathfrak{G}$ acts on an object $\mathcal{A}$ in $\mathcal{VA}_{\mathcal{L}}$ as $\mathfrak{G}(\mathcal{A})=(HOM_{\mathcal{VA}_{\mathcal{L}}}(\mathcal{A},\mathcal{L}),\tau_1,\tau_2,\alpha_{\mathcal{A}})$, where $\alpha_{\mathcal{A}}$ is a mapping from $\mathfrak{S}_{\mathcal{L}}$ to $\Lambda_{HOM_{\mathcal{VA}_{\mathcal{L}}}(\mathcal{A},\mathcal{L})}$ which is defined by $\alpha_{\mathcal{A}}(\mathcal{L}^*)=HOM_{\mathcal{VA}_{\mathcal{L}}}(\mathcal{A},\mathcal{L}^*)$, $\mathcal{L}^*\in\mathfrak{S}_{\mathcal{L}}$.
\item $\mathfrak{G}$ acts on an arrow $\psi:\mathcal{A}\to\mathcal{A}^*$ in $\mathcal{VA}_{\mathcal{L}}$ as follows: $\mathfrak{G}(\psi):\mathfrak{G}(\mathcal{A}^*)\to\mathfrak{G}(\mathcal{A})$ is defined by $\mathfrak{G}(\psi)(\phi)=\phi\circ\psi$, $\phi\in\mathfrak{G}(\mathcal{A}^*)$.
\end{enumerate}
\end{defn}
In \cite{das2021bitopological}, the following duality result is proved for $\mathcal{L}$-$\mathcal{VL}$-algebras:
\begin{thm}
\label{DRVL}
The categories $\mathcal{VA}_{\mathcal{L}}$ and $PBS_{\mathcal{L}}$ are dually equivalent.
\end{thm}
\section{Bitopological duality for Fitting’s many-valued modal logic}
\label{BDFML}
This section extends modal natural duality framework developed by Maruyama in \cite{maruyama2012natural} to the bitopological language. On the semantic side we work with the category $PRBS_{\mathcal{L}}$ of $\mathcal{L}$-valued relational pairwise Boolean spaces. We establish $\mathcal{MA}_{\mathcal{L}}\simeq PRBS_{\mathcal{L}}$. When $\mathcal{L}=\mathbf{2}$, the two-element Boolean algebra, our duality shows the J{\'o}nsson-Tarski duality for modal algebras.\\
Let $\mathcal{R}$ be a relation on $P$ and $C\subseteq P$. We define \\
$[\mathcal{R}]C=\{p\in P:\mathcal{R}[p]\subseteq C\}$ and $\langle \mathcal{R}\rangle C=\{p\in P:\mathcal{R}[p]\cap C\neq\emptyset\}$.
\subsection{Category}
\begin{defn}
	\label{PRBSL}
	We define a category $PRBS_{\mathcal{L}}$ as follows:
	\begin{enumerate}[(1)]
		\item Objects: An object in $PRBS_{\mathcal{L}}$ is a triple  $(P,\alpha_P,\mathcal{R})$ such that $(P,\alpha_P)$ is an object in $PBS_{\mathcal{L}}$ and $\mathcal{R}$ is a binary relation on $P$ that satisfies the following conditions:
		\begin{enumerate}[(i)]
			\item for each $p$ in $P$, $\mathcal{R}[p]$ is a pairwise compact subset of $P$;
			\item $\forall \mathcal{C}\in\beta_1$, $[\mathcal{R}]\mathcal{C}, \langle \mathcal{R}\rangle \mathcal{C}\in \beta_1$;
			\item for any $\mathcal{L'} \in\mathfrak{S}_{\mathcal{L}}$, if $m\in\alpha_P(\mathcal{L'})$ then $\mathcal{R}[m]\subseteq\alpha_P(\mathcal{L'})$.
		\end{enumerate}
		\item Arrows: An arrow $f:(P,\alpha_P,\mathcal{R})\to (P',\alpha_{P'},\mathcal{R}')$ in $PRBS_{\mathcal{L}}$ is an arrow	in $PBS_{\mathcal{L}}$ which additionally satisfies the following conditions:
		\begin{enumerate}[(i)]
			\item if $p_1\mathcal{R}p_2$ then $f(p_1)\mathcal{R}'f(p_2)$;
			\item if $f(p)\mathcal{R}'p'$ then $\exists p^*\in P$ such that $p\mathcal{R}p^*$ and $f(p^*)=p'$.
		\end{enumerate}
	\end{enumerate}
\end{defn}
\begin{note}
	We see that $[\mathcal{R}]Q^c=(\langle \mathcal{R}\rangle Q)^c$, and $\langle \mathcal{R}\rangle Q^c=([\mathcal{R}]Q)^c$. Since $\beta_2=\{Q^c:Q\in\beta_1\}$, hence if the relation $\mathcal{R}$ satisfies condition $(ii)$ that is given in the object part of Definition \ref{PRBSL}, then $[\mathcal{R}]Q, \langle \mathcal{R}\rangle Q\in\beta_2$, $\forall Q\in\beta_2$.
\end{note}
To prove that the category $\mathcal{MA}_{\mathcal{L}}$ is dually equivalent to the category $PRBS_{\mathcal{L}}$, we define two functors $\mathcal{F}$ and $\mathcal{G}$ in the next sequel.
\subsection{Functors}
\begin{defn}
	\label{FUNC3}
	We define a functor $\mathcal{G}:\mathcal{MA}_{\mathcal{L}}\to PRBS_{\mathcal{L}}$.
	\begin{enumerate}[(i)] 
		\item For an object $(\mathcal{A},\Box)$ in $\mathcal{MA}_{\mathcal{L}}$, define  $\mathcal{G}(\mathcal{A})=(HOM_{\mathcal{VA}_{\mathcal{L}}}(\mathcal{A},\mathcal{L}),\tau_1,\tau_2,\alpha_{\mathcal{A}},\mathcal{R}_{\Box})$, where $\alpha_{\mathcal{A}}:\mathfrak{S}_{\mathcal{L}} \to \Lambda_{HOM_{\mathcal{VA}_{\mathcal{L}}}(\mathcal{A},\mathcal{L})}$  is a mapping which is defined by $\alpha_{\mathcal{A}}(\mathcal{L}_1)=HOM_{\mathcal{VA}_{\mathcal{L}}}(\mathcal{A},\mathcal{L}_1)$, and $\mathcal{R}_{\Box}$ is a binary relation on  $HOM_{\mathcal{VA}_{\mathcal{L}}}(\mathcal{A},\mathcal{L})$ that is described in Definition \ref{BR}.
		\item $\mathcal{G}$ acts on an arrow $\psi:\mathcal{A}_1\to\mathcal{A}_2$ in $\mathcal{MA}_{\mathcal{L}}$ as follows: \\
		Define $\mathcal{G}(\psi):\mathcal{G}(\mathcal{A}_2)\to\mathcal{G}(\mathcal{A}_1)$ by $\mathcal{G}(\psi)(\phi)=\phi\circ\psi$, where $\phi\in HOM_{\mathcal{VA}_{\mathcal{L}}}(\mathcal{A}_2,\mathcal{L})$.
	\end{enumerate}
\end{defn}
Lemma \ref{WDG1} and Lemma \ref{WDG2} demonstrate the well-definedness of $\mathcal{G}$.
\begin{lem}
	\label{WDG1}
	For an object  $(\mathcal{A},\Box)$ in $\mathcal{MA}_{\mathcal{L}}$, $\mathcal{G}(\mathcal{A})$ is an object in $PRBS_{\mathcal{L}}$.
	\begin{proof}
		Definition \ref{FUNC2} shows that $(HOM_{\mathcal{VA}_{\mathcal{L}}}(\mathcal{A},\mathcal{L}),\tau_1,\tau_2,\alpha_{\mathcal{A}})$ is an object in $PBS_{\mathcal{L}}$. Thus, it suffices to demonstrate $\mathcal{R}_{\Box}$ satisfies the requirements given in the object section of Definition \ref{PRBSL}. First, we ensure that for any $\mathcal{W}\in HOM_{\mathcal{VA}_{\mathcal{L}}}(\mathcal{A},\mathcal{L})$, $\mathcal{R}_{\Box}[\mathcal{W}]\in\delta_1\cup\delta_2$. Let $\mathcal{U}\not\in \mathcal{R}_{\Box}[\mathcal{W}]$. Then by Definition \ref{BR}, there is an element $a\in\mathcal{A}$ such that there is $L_1\in\mathcal{L}$, for which $\mathcal{W}(\Box a)\geq L_1$ but $\mathcal{U}(a)\not\geq L_1$. It follows that $\mathcal{U}\in\langle \neg U_{L_1}(a)\rangle\in\tau_2$ and $\mathcal{R}_{\Box}[\mathcal{W}]\cap \langle \neg U_{L_1}(a)\rangle=\emptyset$ i.e., $\langle \neg U_{L_1}(a)\rangle\subseteq (\mathcal{R}_{\Box}[\mathcal{W}])^c$. Hence, $\mathcal{U}\not\in\overline{\mathcal{R}_{\Box}[\mathcal{W}]}^{\tau_2}$, where $\overline{\mathcal{R}_{\Box}[\mathcal{W}]}^{\tau_2}$ denotes the closure of $\mathcal{R}_{\Box}[\mathcal{W}]$ in $(HOM_{\mathcal{VA}_{\mathcal{L}}}(\mathcal{A},\mathcal{L}),\tau_2)$ . Equivalently, we have $\overline{\mathcal{R}_{\Box}[\mathcal{W}]}^{\tau_2}\subset\mathcal{R}_{\Box}[\mathcal{W}]$. Therefore, $\mathcal{R}_{\Box}[\mathcal{W}]$ is $\tau_2$-closed. Since $(HOM_{\mathcal{VA}_{\mathcal{L}}}(\mathcal{A},\mathcal{L}),\tau_1,\tau_2)$ is pairwise compact, by Proposition \ref{PROP2}, we have $\mathcal{R}_{\Box}[\mathcal{W}]$ is pairwise compact.\\
		Now we verify the condition $(ii)$ in the object part of Definition \ref{PRBSL}.
		Since $\{\langle a\rangle:a\in\mathcal{A}\}\in\beta_1$ and $\{\langle T_1(a)\rightarrow 0\rangle :a\in\mathcal{A}\}\in\beta_2$ are the basis for the topologies $\tau_1$ and $\tau_2$, respectively, so we show that for each $a\in\mathcal{A}$, $\langle \mathcal{R}_{\Box}\rangle(\langle a\rangle)\in\beta_1$ and $[\mathcal{R}_{\Box}]\langle a\rangle\in\beta_1$. We see that
		\begin{equation*}
			\begin{split}
				\langle \mathcal{R}_{\Box}\rangle\langle a\rangle &=\{\mathcal{W}\in HOM_{\mathcal{VA}_{\mathcal{L}}}(\mathcal{A},\mathcal{L}):\mathcal{R}_{\Box}[\mathcal{W}]\cap\langle a\rangle\neq\emptyset\}\\
				&=([\mathcal{R}_{\Box}]\langle T_1(a)\rightarrow 0\rangle)^c\\
				&=\{\mathcal{W}\in HOM_{\mathcal{VA}_{\mathcal{L}}}(\mathcal{A},\mathcal{L}):\mathcal{R}_{\Box}[\mathcal{W}]\not\subset\langle T_1(a)\rightarrow 0\rangle \}
			\end{split}
		\end{equation*}
	We show that $([\mathcal{R}_{\Box}]\langle T_1(a)\rightarrow 0\rangle)^c$ is $\tau_1$-open and $\tau_2$-closed. Let $\mathcal{U}\in ([\mathcal{R}_{\Box}]\langle T_1(a)\rightarrow 0\rangle)^c$. Then $\mathcal{R}_{\Box}[\mathcal{U}]\not\subset\langle T_1(a)\rightarrow 0\rangle$. It is easy to see that $\exists$ $\tau_1$-open set $\langle \Box T_1(a)\rangle$ such that $\mathcal{U}\in \langle \Box T_1(a)\rangle$. Let $\mathcal{E}\in \langle \Box T_1(a)\rangle$. Then $\mathcal{E}(\Box T_1(a))=1$. Using the Kripke condition we have $1=\mathcal{E}(\Box T_1(a))=\bigwedge\{\mathcal{U}(T_1(a)):\mathcal{E}\mathcal{R}_{\Box}\mathcal{U}\}$. According to Lemma \ref{TF2}, $\mathcal{U}(T_1(a))$ is either $0$ or $1$. Henceforth, for all $\mathcal{U}\in HOM_{\mathcal{VA}_{\mathcal{L}}}(\mathcal{A},\mathcal{L})$ with $\mathcal{E}\mathcal{R}_{\Box}\mathcal{U}$ we have $\mathcal{U}(T_1(a))=1$. As a result, $\mathcal{R}_{\Box}[\mathcal{E}]\not\subset\langle T_1(a)\rightarrow 0\rangle$ i.e., $\mathcal{E}\in ([\mathcal{R}_{\Box}]\langle T_1(a)\rightarrow 0\rangle)^c$. Henceforth, $\mathcal{U}\in\langle \Box T_1(a)\rangle\subset ([\mathcal{R}_{\Box}]\langle T_1(a)\rightarrow 0\rangle)^c$. Therefore, $[\mathcal{R}_{\Box}]\langle T_1(a)\rightarrow 0\rangle)^c$ is $\tau_1$-open i.e., $\langle \mathcal{R}_{\Box}\rangle\langle a\rangle$ is $\tau_1$-open. \\
	Let $\mathcal{W}\in (\langle \mathcal{R}_{\Box}\rangle\langle a\rangle)^c$. Then $\mathcal{R}_{\Box}[\mathcal{W}]\cap\langle a\rangle=\emptyset$. It is easy to see that there is $\tau_1$-open set $\langle \Box(T_1(a)\rightarrow0)\rangle$ such that $\mathcal{W}\in\langle \Box(T_1(a)\rightarrow 0)\rangle$. Also, by applying the Kripke condition, we have $\langle \Box(T_1(a)\rightarrow 0)\rangle\subset (\langle \mathcal{R}_{\Box}\rangle\langle a\rangle)^c$. Therefore, $\mathcal{W}\in\langle \Box(T_1(a)\rightarrow 0)\rangle \subset (\langle \mathcal{R}_{\Box}\rangle\langle a\rangle)^c$. It shows that $(\langle \mathcal{R}_{\Box}\rangle\langle a\rangle)^c$ is $\tau_1$-open i.e., $\langle \mathcal{R}_{\Box}\rangle\langle a\rangle$ is $\tau_1$-closed. It follows from Proposition \ref{PROP2} that $\langle \mathcal{R}_{\Box}\rangle\langle a \rangle$ is pairwise compact. Since the topological space $(HOM_{\mathcal{VA}_{\mathcal{L}}}(\mathcal{A},\mathcal{L}),\tau_2)$ with basis $\{\langle T_1(a)\rightarrow 0\rangle: a\in\mathcal{A}\}$ is a Hausdorff space, so $\langle \mathcal{R}_{\Box}\rangle\langle a\rangle$ is $\tau_2$-closed. Hence, $\langle\mathcal{R}_{\Box}\rangle\langle a\rangle\in\beta_1$.\\
	Next, we show that $[\mathcal{R}_{\Box}]\langle a\rangle\in\beta_1$. We see that
	\begin{equation*}
		\begin{split}
			[\mathcal{R}_{\Box}]\langle a\rangle &=\{\mathcal{W}\in HOM_{\mathcal{VA}_{\mathcal{L}}}(\mathcal{A},\mathcal{L}):\mathcal{R}_{\Box}[\mathcal{W}]\subseteq \langle a\rangle\}\\
			&=(\langle \mathcal{R}_{\Box}\rangle\langle T_1(a)\rightarrow 0\rangle)^c
		\end{split}
	\end{equation*}
		We claim that $\langle \mathcal{R}_{\Box}\rangle\langle T_1(a)\rightarrow 0\rangle=\langle \Box T_1(a)\rightarrow 0\rangle$. Let $\mathcal{W}\in\langle \Box T_1(a)\rightarrow 0\rangle$. Then $\mathcal{W}(\Box T_1(a)\rightarrow 0)=1$. Hence, $\mathcal{W}(\Box T_1(a))=0$. Using the Kripke condition, we have, $0=\mathcal{W}(\Box T_1(a))=\bigwedge\{\mathcal{U}(T_1(a)):\mathcal{W}\mathcal{R}_{\Box}\mathcal{U}\}$. Since $\mathcal{U}(T_1(a))=0 \text{ or } 1$, hence $\exists$ $\mathcal{U}\in HOM_{\mathcal{VA}_{\mathcal{L}}}(\mathcal{A},\mathcal{L})$ with $\mathcal{W}\mathcal{R}_{\Box}\mathcal{U}$ such that $\mathcal{U}(T_1(a))=0$. Then $\mathcal{U}\in\langle T_1(a)\rightarrow 0\rangle$. Therefore, $\mathcal{R}_{\Box}[\mathcal{W}]\cap\langle T_1(a)\rightarrow 0\rangle\neq\emptyset$. Thus, $\mathcal{W}\in\langle\mathcal{R}_{\Box}\rangle\langle T_1(a)\rightarrow 0\rangle$. Similarly, by employing the Kripke condition, we can show that if $\mathcal{W}\in\langle\mathcal{R}_{\Box}\rangle\langle T_1(a)\rightarrow 0\rangle$ then $\mathcal{W}\in\langle\Box T_1(a)\rightarrow 0\rangle$. Since $\langle\Box T_1(a)\rightarrow 0\rangle\in\beta_2$, we have $\langle\mathcal{R}_{\Box}\rangle\langle T_1(a)\rightarrow 0\rangle\in\beta_2$. As a result, $[\mathcal{R}_{\Box}]\langle a\rangle\in\beta_1$.\\
	Finally, we demonstrate that $\mathcal{G}(\mathcal{A})$ meets condition $(iii)$ in the object part of Definition \ref{PRBSL}. Let $u\in\alpha_{\mathcal{A}}(\mathcal{L}')=HOM_{\mathcal{VA}_{\mathcal{L}}}(\mathcal{A},\mathcal{L}')$. Suppose $\mathcal{R}_{\Box}[u]\not\subset \alpha_{\mathcal{A}}(\mathcal{L}')$. Then $\exists v\in\mathcal{R}_{\Box}[u]$ such that $v\notin\alpha_{\mathcal{A}}(\mathcal{L}')$. Hence, $\exists a^*\in\mathcal{A}$ such that $v(a^*)\notin\mathcal{L}'$. Let $v(a^*)=\ell^*$. Now for any element $\psi\in\alpha_{\mathcal{A}}(\mathcal{L}')$,
\[ \psi(T_{\ell^*}(a^*)\rightarrow a^*)=  \left\{
\begin{array}{ll}
\ell^* & \text{ if }\psi(a^*)=\ell^*\\
1 & \text{ if } \psi(a^*)\neq\ell^* \\
\end{array} 
\right. \]
Using Kripke condition, we have $u(\Box(T_{\ell^*}(a^*)\rightarrow a^*))=\bigwedge\{\psi(T_{\ell^*}(a^*)\rightarrow a^*):\psi\in\mathcal{R}_{\Box}[u]\}$. This shows that $u(\Box(T_{\ell^*}(a^*)\rightarrow a^*))=\ell^*\notin\mathcal{L}'$. But this contradicts the fact that $u\in\alpha_{\mathcal{A}}(\mathcal{L}')$. As a result, $\mathcal{G}(\mathcal{A})$ satisfies condition $(iii)$.
	\end{proof}
\end{lem}
\begin{lem}
\label{WDG2}
Let $(\mathcal{A}_1,\Box_1)$, $(\mathcal{A}_2,\Box_2)$ be  the objects in $\mathcal{MA}_{\mathcal{L}}$ and $\psi:\mathcal{A}_1\to\mathcal{A}_2$ be an arrow in $\mathcal{MA}_{\mathcal{L}}$. Then, $\mathcal{G}(\psi)$ is an arrow in $PRBS_{\mathcal{L}}$.
\begin{proof}
Here $\mathcal{G}(\psi):\mathcal{G}(\mathcal{A}_2)\to\mathcal{G}(\mathcal{A}_1)$ is defined by $\mathcal{G}(\psi)(\phi)=\phi\circ\psi$, $\phi\in HOM_{\mathcal{VA}_{\mathcal{L}}}(\mathcal{A}_2,\mathcal{L})$. It follows from Definition \ref{FUNC2} that $\mathcal{G}(\psi)$ is an arrow in $PBS_{\mathcal{L}}$. Therefore, it is still necessary to demonstrate that $\mathcal{G}(\psi)$ satisfies conditions $(i)$ and $(ii)$ listed in the arrow portion of Definition \ref{PRBSL}. We first check condition $(i)$. Let $v_1\mathcal{R}_{\Box_2}v_2$, where $v_1, v_2\in\mathcal{G}(\mathcal{A}_2)$. We are to show that $\mathcal{G}(\psi)(v_1)\mathcal{R}_{\Box_1}\mathcal{G}(\psi)(v_2)$. Now, if $v_1\circ\psi(\Box_1 a_1)\geq\ell$ for $a_1\in\mathcal{A}_1$ and $\ell\in\mathcal{L}$, then we have $v_1(\Box_2\psi(a_1))\geq\ell$. As $v_1\mathcal{R}_{\Box_2}v_2$, so we get $v_2(\psi(a_1))\geq\ell$. Hence, $\mathcal{G}(\psi)(v_1)\mathcal{R}_{\Box_1}\mathcal{G}(\psi)(v_2)$. We then check condition $(ii)$, which is mentioned in the arrow part of Definition \ref{PRBSL}. This is equivalent to verifying $\mathcal{R}_{\Box_1}[\mathcal{G}(\psi)(v_1)]=\mathcal{G}(\psi)(\mathcal{R}_{\Box_2}[v_1])$. Let $\mathcal{W}\in\mathcal{R}_{\Box_1}[v_1\circ\psi]$, where $\mathcal{W}\in HOM_{\mathcal{VA}_{\mathcal{L}}}(\mathcal{A}_1,\mathcal{L})$. Then $(v_1\circ\psi)\mathcal{R}_{\Box_1}\mathcal{W}$. Suppose $\mathcal{W}\notin\mathcal{G}(\psi)(\mathcal{R}_{\Box_2}[v_1])$. Then $\mathcal{W}\neq\mathcal{G}(\psi)(v^*)$, $\forall v^*\in HOM_{\mathcal{VA}_{\mathcal{L}}}(\mathcal{A}_2,\mathcal{L})$ such that $v_1\mathcal{R}_{\Box_2}v^*$. As $(HOM_{\mathcal{VA}_{\mathcal{L}}}(\mathcal{A}_1,\mathcal{L}),\tau_1,\tau_2)$ is a pairwise Hausdorff space, so we can consider $\mathcal{W}\in\langle a_1\rangle$ and $\mathcal{G}(\psi)(v^*)=v^*\circ\psi\in\langle T_1(a_1)\rightarrow 0\rangle$. Since $\mathcal{W}\in\mathcal{R}_{\Box_1}[\mathcal{G}(\psi)(v_1)]$ and $\mathcal{W}(a_1)=1$, we have $\mathcal{G}(\psi)(v_1)(\Box_1 a_1)=1$ i.e., $(v_1\circ\psi)(\Box_1 a_1)=1$. Since $v_1\mathcal{R}_{\Box_2}v^*$, we have $\mathcal{G}(\psi)(v_1)\mathcal{R}_{\Box_1}\mathcal{G}(\psi)(v^*)$ using the condition $(i)$ specified in the arrow part of Definition \ref{PRBSL}. As $\mathcal{G}(\psi)(v_1)(\Box_1 a_1)=1$, Lemma \ref{KM} shows that $\mathcal{G}(\psi)(v^*)(a_1)=1$, i.e., $v^*\circ\psi\in\langle a_1\rangle$. This contradicts the fact that $\mathcal{G}(\psi)(v^*)\in\langle T_1(a_1)\rightarrow 0\rangle$. Therefore, $\mathcal{R}_{\Box_1}[\mathcal{G}(\psi)(v_1)]\subseteq\mathcal{G}(\psi)(\mathcal{R}_{\Box_2}[v_1])$. Similarly, we can show the reverse direction.
\end{proof}
\end{lem} 
\begin{defn}
\label{FUNC4}
We define a functor $\mathcal{F}:PRBS_{\mathcal{L}}\to\mathcal{MA}_{\mathcal{L}}$.
			\begin{enumerate}[(i)]
				\item Define $\mathcal{F}(P,\alpha_P,\mathcal{R})=(HOM_{{PBS}_{\mathcal{L}}}((P,\alpha_P),(\mathcal{L},\alpha_{\mathcal{L}})),\wedge,\vee,\rightarrow,T_{\ell}(\ell\in\mathcal{L}),0,1,\Box_{\mathcal{R}})$ for an object $(P,\alpha_P,\mathcal{R})$ in $PRBS_{\mathcal{L}}$. Definition \ref{MO} describes the modal operation $\Box_{\mathcal{R}}$. Here $\wedge,\vee,\rightarrow, T_{\ell}$ are pointwise operations defined on the set $HOM_{{PBS}_{\mathcal{L}}}((P,\alpha_P),(\mathcal{L},\alpha_{\mathcal{L}}))$.
				\item Let $\psi:(P_1,\alpha_{P_1},\mathcal{R}_1)\to (P_2,\alpha_{P_2},\mathcal{R}_2)$ be an arrow in $PRBS_{\mathcal{L}}$. Define $\mathcal{F}(\psi):\mathcal{F}(P_2,\alpha_{P_2},\mathcal{R}_2)\to\mathcal{F}(P_1,\alpha_{P_1},\mathcal{R}_1)$ by $\mathcal{F}(\psi)(\phi)=\phi\circ\psi$ for $\phi\in \mathcal{F}(P_2,\alpha_{P_2},\mathcal{R}_2)$.
			\end{enumerate}
\end{defn}
\begin{note}
If $\psi,\phi: (P,\tau_1^P,\tau_2^P,\alpha_P)\to(\mathcal{L},\tau,\tau,\alpha_{\mathcal{L}})$ are pairwise continuous maps then $\psi\wedge\phi,\psi\vee\phi,\psi\rightarrow\phi, T_{\ell}(\psi)$ are pairwise continuous maps. Thus,  $(HOM_{{PBS}_{\mathcal{L}}}((P,\alpha_P),(\mathcal{L},\alpha_{\mathcal{L}})),\wedge,\vee,\rightarrow,T_{\ell}(\ell\in\mathcal{L}),0,1)$ is a $\mathcal{L}$-$\mathcal{VL}$-algebra.
\end{note}
Lemmas \ref{WDF1} and \ref{WDF2} demonstrate that the functor $\mathcal{F}$ is well-defined. 
\begin{lem}
\label{WDF1}
Let $(P,\alpha_P,\mathcal{R})$ be an object in $PRBS_{\mathcal{L}}$. Then, $\mathcal{F}(P,\alpha_P,\mathcal{R})$ is an object in $\mathcal{MA}_{\mathcal{L}}$. 
\begin{proof}
It is clear from Definition \ref{FUNC1} that $\mathcal{F}(P,\alpha_P)$ is an object in $\mathcal{VA}_{\mathcal{L}}$. We shall now make sure that the modal operation $\Box_{\mathcal{R}}$ on $\mathcal{F}(P,\alpha_P,\mathcal{R})$ is well-defined. Let $\eta\in\mathcal{F}(P,\alpha_P,\mathcal{R})$.  We then verify $\Box_{\mathcal{R}}\eta\in\mathcal{F}(P,\alpha_P,\mathcal{R})$. For any $\ell\in\mathcal{L}$, 
\begin{equation*}
\begin{split}
(\Box_{\mathcal{R}}\eta)^{-1}(\{\ell\})&=\{p\in P:\bigwedge\{\eta(p'):p'\in\mathcal{R}[p]=\ell\}\\
&=\langle\mathcal{R}\rangle((T_{\ell}(\eta))^{-1}(\{1\}))\cap(\langle\mathcal{R}\rangle((U_{\ell}(\eta))^{-1}(\{0\})))^c
\end{split}
\end{equation*}
As both $T_{\ell}(\eta)$ and $U_{\ell}(\eta)$ are pairwise continuous maps, henceforth $(T_{\ell}(\eta))^{-1}(\{1\})\in\beta_1^P\cap\beta_2^P$ and $(U_{\ell}(\eta))^{-1}(\{0\})\in\beta_1^P\cap\beta_2^P$, where $\beta_1^P=\tau_1^P\cap\delta_2^P$ and $\beta_2^P=\tau_2^P\cap\delta_1^P$. Therefore, $(\Box_{\mathcal{R}}\eta)^{-1}(\{\ell\})\in\tau_1^P$. Also, $(\Box_{\mathcal{R}}\eta)^{-1}(\{\ell\})\in\tau_2^P$. As a result, $\Box_{\mathcal{R}}\eta$ is a pairwise continuous map from $P$ to $\mathcal{L}$. Furthermore, by applying condition $(iii)$ that is stated in the object part of Definition \ref{PRBSL}, we see that for any subalgebra $\mathcal{M}\in\mathfrak{S}_{\mathcal{L}}$, and if $m\in\alpha_{\mathcal{L}}(\mathcal{M})$ then $(\Box_{\mathcal{R}}\eta)(m)=\bigwedge\{\eta(m'):m'\in\mathcal{R}[m]\}\in\alpha_{\mathcal{L}}(\mathcal{M})$. Thus $\Box_{\mathcal{R}}\eta$ is a subspace preserving map. Hence, 
$\Box_{\mathcal{R}}\eta\in\mathcal{F}(P,\alpha_P,\mathcal{R})$.
\end{proof}
\end{lem}
\begin{lem}
\label{WDF2}
Let $\psi:(P_1,\alpha_{P_1},\mathcal{R}_1)\to (P_2,\alpha_{P_2},\mathcal{R}_2)$ be an arrow in $PRBS_{\mathcal{L}}$. Then, $\mathcal{F}(\psi)$ is an arrow in $\mathcal{MA}_{\mathcal{L}}$. 
\begin{proof}
According to Definition \ref{FUNC1}, $\mathcal{F}(\psi)$ is an arrow in $\mathcal{VA}_{\mathcal{L}}$. Therefore, it is sufficient to demonstrate that $\mathcal{F}(\psi)(\Box_{\mathcal{R}_2}\phi_2)=\Box_{\mathcal{R}_1}(\mathcal{F}(\psi)\phi_2)$, where $\phi_2\in HOM_{{PBS}_{\mathcal{L}}}((P_2,\alpha_{P_2}),(\mathcal{L},\alpha_{\mathcal{L}}))$. For any $p_1\in P_1$, we have $\mathcal{F}(\psi)(\Box_{\mathcal{R}_2}\phi_2)(p_1)=\Box_{\mathcal{R}_2}\phi_2\circ\psi(p_1)=\bigwedge\{\phi_2(p_2):p_2\in\mathcal{R}_2[\psi(p_1)]\}$, and $\Box_{\mathcal{R}_1}(\mathcal{F}(\psi)\phi_2)(p_1)=\Box_{\mathcal{R}_1}(\phi_2\circ\psi)(p_1)=\bigwedge\{\phi_2\circ\psi(p): p\in\mathcal{R}_1[p_1]\}$. As $\psi$ satisfies conditions $(i)$ and $(ii)$ listed in item 2 of Definition \ref{PRBSL}, it is easy to show that $(\mathcal{F}(\psi)(\Box_{\mathcal{R}_2}\phi_2))(p_1)\leq \Box_{\mathcal{R}_1}(\mathcal{F}(\psi)\phi_2)(p_1)$ and $\Box_{\mathcal{R}_1}(\mathcal{F}(\psi)\phi_2)(p_1)\leq (\mathcal{F}(\psi)(\Box_{\mathcal{R}_2}\phi_2))(p_1)$. As a result, $\mathcal{F}(\psi)(\Box_{\mathcal{R}_2}\phi_2)=\Box_{\mathcal{R}_1}(\mathcal{F}(\psi)\phi_2)$.
\end{proof}
\end{lem}
\subsection{Bitopological Duality for $\mathcal{L}$-$\mathcal{ML}$ Algebras}
In this subsection, we develop bitopological duality for algebras of Fitting’s Heyting-valued modal logic. 
\begin{thm}
\label{BTHM1}
	Let $\mathcal{A}$ be a $\mathcal{L}$-$\mathcal{ML}$ algebra. Then $\mathcal{A}$ is isomorphic to $\mathcal{F}\circ\mathcal{G}(\mathcal{A})$ in $\mathcal{MA}_{\mathcal{L}}$.
\begin{proof}
We define $\gamma^{\mathcal{A}}:\mathcal{A}\to\mathcal{F}\circ\mathcal{G}(\mathcal{A})$ by $\gamma^{\mathcal{A}}(a)(g)=g(a)$, where $a\in\mathcal{A}$ and $g\in HOM_{\mathcal{VA}_{\mathcal{L}}}(\mathcal{A},\mathcal{L})$. It is known from Theorem \ref{DRVL} that $\gamma^{\mathcal{A}}$ is an isomorphism in the category $\mathcal{VA}_{\mathcal{L}}$. The only thing left to prove is that $\gamma^{\mathcal{A}}$ preserves the modal operation
$\Box$, i.e., $\gamma^{\mathcal{A}}(\Box a)=\Box_{\mathcal{R}_{\Box}}\gamma^{\mathcal{A}}(a)$, $a\in\mathcal{A}$. Let $g\in\mathcal{G}(\mathcal{A})$. Then
\begin{equation*}
\begin{split}
(\Box_{\mathcal{R}_{\Box}}\gamma^{\mathcal{A}}(a))(g)&=\bigwedge\{\gamma^{\mathcal{A}}(g^*):g^*\in\mathcal{R}_{\Box}[g]\}\\
&=\bigwedge\{g^*(a):g^*\in\mathcal{R}_{\Box}[g]\}\\
&=g(\Box a) \text{ (by Lemma \ref{KM})}\\
&=\gamma^{\mathcal{A}}(\Box a)(g)
\end{split}
\end{equation*}
Hence the result follows.
\end{proof}
\end{thm}
\begin{thm}
\label{BTHM2}
Consider an object $(P,\alpha_{P},\mathcal{R})$ in $PRBS_{\mathcal{L}}$. Then, $(P,\alpha_{P},\mathcal{R})$ is isomorphic to $\mathcal{G}\circ\mathcal{F}(P,\alpha_P,\mathcal{R})$ in the category $PRBS_{\mathcal{L}}$.
\begin{proof}
Define $\zeta_{(P,\alpha_P,\mathcal{R})}:(P,\alpha_P,\mathcal{R})\to\mathcal{G}\circ\mathcal{F}(P,\alpha_P,\mathcal{R})$ by $\zeta_{(P,\alpha_P,\mathcal{R})}(p)(\psi)=\psi(p)$, where $p\in P$ and $\psi\in HOM_{{PBS}_{\mathcal{L}}}((P,\alpha_P),(\mathcal{L},\alpha_{\mathcal{L}}))$. Theorem \ref{DRVL} shows that $\zeta_{(P,\alpha_P,\mathcal{R})}$ is a bi-homeomorphism in the category $PBS_{\mathcal{L}}$. We show that $\zeta_{(P,\alpha_P,\mathcal{R})}$ and $\zeta^{-1}_{(P,\alpha_P,\mathcal{R})}$ satisfy the conditions given in item 2 of Definition \ref{PRBSL}. We claim that for any $p,p'\in P$, $p'\in\mathcal{R}[p]\iff \zeta_{(P,\alpha_P,\mathcal{R})}(p')\in\mathcal{R}_{\Box_{\mathcal{R}}}[\zeta_{(P,\alpha_P,\mathcal{R})}(p)]$. Let $p'\in\mathcal{R}[p]$. Suppose $\zeta_{(P,\alpha_P,\mathcal{R})}(p)(\Box_{\mathcal{R}}\psi)\geq\ell$, where $\ell\in\mathcal{L}$ and $\psi\in HOM_{{PBS}_{\mathcal{L}}}((P,\alpha_P),(\mathcal{L},\alpha_{\mathcal{L}}))$. Then $\zeta_{(P,\alpha_P,\mathcal{R})}(p)(\Box_{\mathcal{R}}\psi)=(\Box_{\mathcal{R}}\psi)(p)=\bigwedge\{\psi(p^*):p*\in\mathcal{R}[p]\}$. Since $p'\in\mathcal{R}[p]$ and $\zeta_{(P,\alpha_P,\mathcal{R})}(p)(\Box_{\mathcal{R}}\psi)\geq\ell$, we have $\zeta_{(P,\alpha_P,\mathcal{R})}(p')(\psi)\geq\ell$. Hence, $\zeta_{(P,\alpha_P,\mathcal{R})}(p)\mathcal{R}_{\Box_{\mathcal{R}}} \zeta_{(P,\alpha_P,\mathcal{R})}(p')$, i.e., $\zeta_{(P,\alpha_P,\mathcal{R})}(p')\in \mathcal{R}_{\Box_{\mathcal{R}}}[\zeta_{(P,\alpha_P,\mathcal{R})}(p)]$. Now we verify if $\zeta_{(P,\alpha_P,\mathcal{R})}(p')\in \mathcal{R}_{\Box_{\mathcal{R}}}[\zeta_{(P,\alpha_P,\mathcal{R})}(p)]$ then $p'\in\mathcal{R}[p]$. We verify its contrapositive statement. Suppose $p'\notin\mathcal{R}[p]$. By Definition \ref{PRBSL}, $\mathcal{R}[p]$ is a pairwise compact subset of pairwise Boolean space $P$. Then It is straightforward to demonstrate that $\mathcal{R}[p]$ is pairwise closed. Therefore we can get a $\tau_1^P$-basis open set $\mathcal{O}\in\beta_1^P$ such that $p'\in\mathcal{O}$ and $\mathcal{O}\subseteq P-\mathcal{R}[p]$, i.e., $\mathcal{O}\cap\mathcal{R}[p]=\emptyset$. Define a mapping $f:P\to\mathcal{L}$ by 
\[ f(p)=  \left\{
\begin{array}{ll}
	0 & \text{ if $p\in\mathcal{O}$}\\
	1 & \text{ if $p\in\mathcal{O}^c$} \\
\end{array} 
\right. \]
Then $f$ is a pairwise continuous map from $(P,\tau_1^P,\tau_2^P)$ to $(\mathcal{L},\tau,\tau)$. As a result, it can be shown that $f\in HOM_{PBS_{\mathcal{L}}}((P,\alpha_P),(\mathcal{L},\alpha_{\mathcal{L}}))$. Now, $\Box_{\mathcal{R}}f(p)=\bigwedge\{f(z):z\in\mathcal{R}[p]\}=1$ and $f(p')=0$. Hence, $\zeta_{(P,\alpha_P,\mathcal{R})}(p)(\Box_{\mathcal{R}}f)=1$ but $\zeta_{(P,\alpha_P,\mathcal{R})}(p')(f)\neq 1$. Therefore, $\zeta_{(P,\alpha_P,\mathcal{R})}(p')\notin\mathcal{R}_{\Box_{\mathcal{R}}}[\zeta_{(P,\alpha_P,\mathcal{R})}(p)]$. Hence, we have for any $p,p'\in P$, $p'\in\mathcal{R}[p]\iff \zeta_{(P,\alpha_P,\mathcal{R})}(p')\in\mathcal{R}_{\Box_{\mathcal{R}}}[\zeta_{(P,\alpha_P,\mathcal{R})}(p)]$. As a result, $\zeta_{(P,\alpha_P,\mathcal{R})}$ and $\zeta^{-1}_{(P,\alpha_P,\mathcal{R})}$ satisfy conditions $(i)$ and $(ii)$ mentioned in item 2 of Definition \ref{PRBSL}.
\end{proof}
\end{thm}
Finally, we obtain the bitopological duality for Fitting’s Heyting-valued modal logic.
\begin{thm}
	\label{DRML}
The categories $\mathcal{MA}_{\mathcal{L}}$ and $PRBS_{\mathcal{L}}$ are dually equivalent.
\begin{proof}
	Let $ID_1$ and $ID_2$ be the identity functors on $\mathcal{MA}_{\mathcal{L}}$  and $PRBS_{\mathcal{L}}$, respectively. This theorem will be proved by defining two natural isomorphisms, $\gamma:ID_1\to\mathcal{F}\circ\mathcal{G}$ and $\zeta:ID_2\to\mathcal{G}\circ\mathcal{F}$. For an object $\mathcal{A}$ in $\mathcal{MA}_{\mathcal{L}}$ define $\gamma^{\mathcal{A}}:\mathcal{A}\to \mathcal{F}\circ\mathcal{G}(\mathcal{A})$ by $\gamma^{\mathcal{A}}(a)(g)=g(a)$, where $a\in\mathcal{A}$ and $g\in\mathcal{G}(\mathcal{A})$. For an object $(P,\alpha_P,\mathcal{R})$ in $PRBS_{\mathcal{L}}$ define $\zeta_{(P,\alpha_P,\mathcal{R})}:(P,\alpha_P,\mathcal{R})\to \mathcal{G}\circ\mathcal{F}(P,\alpha_P,\mathcal{R})$ by $\zeta_{(P,\alpha_P,\mathcal{R})}(p)(\psi)=\psi(p)$, where $p\in P$ and $\psi\in HOM_{PBS_{\mathcal{L}}}((P,\alpha_P),(\mathcal{L},\alpha_{\mathcal{L}}))$. Then it can be shown that $\gamma$ and $\zeta$ are natural transformations. According to Theorems \ref{BTHM1} and \ref{BTHM2}, $\gamma$ and $\zeta$ are natural isomorphisms.
\end{proof}
\end{thm}
%\begin{rem}
	
%\end{rem}
In the next section, we develop a coalgebraic duality theorem for algebras
of Fitting’s Heyting valued modal logic.
\section{Coalgebraic Duality Theorem for Fitting's multi-valued logic}\label{COALGD}
The aim of this section is to adapt the bi-Vietoris construction of Lauridsen on pairwise Stone spaces \cite{lauridsen2015bitopological} to our $\mathcal{L}$-valued pairwise Boolean spaces. We thereby define an endofunctor $V_{\mathcal{L}}^{bi}$ on the category $PBS_{\mathcal{L}}$ by incorporating the structure map, and we show that $V_{\mathcal{L}}^{bi}$ preserves the $\mathcal{L}$-valued structure. After that, we exhibit that the category $PRBS_{\mathcal{L}}$ is isomorphic to the category $COALG(V_{\mathcal{L}}^{bi})$ of coalgebras for the endofunctor $V_{\mathcal{L}}^{bi}$, and combine this identity with the bitopological duality developed in Section \ref{BDFML} to establish a coalgebraic duality $\mathcal{MA}_{\mathcal{L}}\simeq COALG(V_{\mathcal{L}}^{bi})^{op}$ from which soundness and completeness of Fitting's $\mathcal{L}$-valued modal logic with respect to $V_{\mathcal{L}}^{bi}$-coalgebras follow.\\
In this framework, our assumption over $\mathcal{L}$, the algebra of truth values, is same as in previuos sections. We work with Fitting's-style $\mathcal{L}$-valued modal logic that is modified by Maruyama in his work \cite{maruyama2009algebraic}. The underlying algebra of the $\mathcal{L}$-$\mathcal{VL}$-algebra is Heyting, and hence pseudo-complement exists.
However, the modal signature in this framework is negation-free, i.e., neither Boolean nor involutive negation. The modal operator $\Box$ preserves finite meets and $\top$ element, and commutes with each $U_{\ell}$, $\ell\in\mathcal{L}$.\\
We refer the reader to \cite{adamek2005introduction} for an overview of coalgebras. Now, let us review the concepts of coalgebra and coalgebra morphisms.
\begin{defn}
A coalgebra for an endofunctor $\mathfrak{T}:\bf{C}\to\bf{C}$ on a category $\bf{C}$, called $\mathfrak{T}$-coalgebra, is defined by a tuple $(C,T)$, where $C$ is an object in $\bf{C}$ and $T:C\to \mathfrak{T}(C)$ is an arrow in $\bf{C}$.
\end{defn}
\begin{defn}
	Let $(C_1,T_1)$ and $(C_2,T_2)$ be any two $\mathfrak{T}$-coalgebras. Then $f:(C_1,T_1)\to (C_2,T_2)$ is said to be a $\mathfrak{T}$-coalgebra morphism if $f:C_1\to C_2$ is an arrow in $\bf{C}$ which satisfies $T_2\circ f=\mathfrak{T}(f)\circ T_1$.
\end{defn}
$\mathfrak{T}$-coalgebras and $\mathfrak{T}$-coalgebra morphisms form a category, denoted by $COALG(\mathfrak{T})$.
\subsection{The structure of the endofunctor $V_{\mathcal{L}}^{bi}$}
To set up the coalgebraic presentation of our bitopological duality for Fitting's $\mathcal{L}$-valued modal logic, we recall the pairwise Vietoris space construction and its basic properties on pairwise Stone spaces from \cite{lauridsen2015bitopological}.\\
The notion of pairwise Vietoris space is given in the following definition:
\begin{defn}\label{PVS}\cite{lauridsen2015bitopological}
		Let $(S,\tau_1^S,\tau_2^S)$ be a pairwise topological space and $\mathcal{K}(S)$ the set of all pairwise closed subsets of $S$. We define $\Box U=\{C\in\mathcal{K}(S):C\subseteq U\}$ and $\Diamond U=\{C\in\mathcal{K}(S):C\cap U\neq\emptyset \}$, $U \subseteq S$. Let $\beta_1^S$ and $\beta_2^S$ be the basis for the topologies $\tau_1^S$ and $\tau_2^S$, respectively. The pairwise Vietoris space $V_P(S)$ of the pairwise topological space $(S,\tau_1^S,\tau_2^S)$ is defined as a pairwise topological space $(\mathcal{K}(S),\tau_1^V,\tau_2^V)$, where $\tau_1^V$ is the topology on $\mathcal{K}(S)$ generated by subbasis $\{\Box U,\Diamond U: U\in \beta_1^S\}$ and the topology $\tau_2^V$ on $\mathcal{K}(S)$ is generated by subbasis $\{\Box U,\Diamond U: U\in \beta_2^S\}$.
\end{defn}
\begin{rem}
In the above definition of $V_P(S)$, $\Box U$ and $\Diamond U$ denote the subbasic open sets. The algebras of Fitting's $\mathcal{L}$-valued modal logic have only one modal operator $\Box$, and its semantics is given via subbasic open set $\Box U$ . We do not use $\Diamond$ as the algebraic signature. Thus, $\Diamond$ acts as a topological generator and doesn't match up with any algebraic modality in our work.
\end{rem}
The following three lemmas (zero-dimensionality, Hausdorffness, compactness of $V_P(S)$, whenever $S$ is a pairwise Boolean space) are of routine in the pairwise Vietoris set-up and can be found in Lauridsen's work \cite{lauridsen2015bitopological}. Together they yield that $V_P(S)$ is a pairwise Boolean space whenever $S$ is a pairwise Boolean space. 
\begin{lem}\label{PZD}
If $(S,\tau_1^S,\tau_2^S)$ is a pairwise Boolean space then $V_P(S)=(\mathcal{K}(S),\tau_1^V,\tau_2^V)$ is pairwise zero-dimensional.
%\begin{proof}
%We shall show that $\beta_1^V=\tau_1^V\cap\delta_2^V$ is a basis for $\tau_1^V$, where $\delta_2^V$ is the set of $\tau_2^V$-closed sets. Let $\mathcal{O}\in\tau_1^V$. Then $\mathcal{O}$ can be expressed as $\mathcal{O}=\displaystyle\bigcup_{\lambda\in\Lambda}(\bigcap_{r=1}^{n_{\lambda}}\Box C_r\cap\bigcap_{t=1}^{m_{\lambda}}\Diamond C_t)$, $C_r, C_t\in\beta_1^S=\tau_1^S\cap\delta_2^S$. To demonstrate that $\beta_1^V$ is a basis for $\tau_1^V$, we have to prove that $\bigcap_{r=1}^{n_{\lambda}}\Box C_r\cap\bigcap_{t=1}^{m_{\lambda}}\Diamond C_t\in\beta_1^V$. Because the finite intersection of the members of $\beta_1^V$ is again in $\beta_1^V$, it is sufficient to establish that for $C\in\beta_1^S$, $\Box C,\Diamond C\in\beta_1^V$. As $\tau_1^V$ is the topology generated by the subbasis $\{\Box C,\Diamond C: C\in\beta_1^S\}$, hence $\Box C,\Diamond C\in\tau_1^V$. Now we see that $(\Box C)^c=\Diamond C^c$ and $(\Diamond C)^c=\Box C^c$. Since $C\in\beta_1^S$, so $C^c\in\beta_2^S$. As a result, $\Box C,\Diamond C\in\delta_2^V$. Henceforth, $\Box C,\Diamond C\in\beta_1^V$. Similarly, it can be shown that $\beta_2^V=\tau_2^V\cap\delta_1^V$, $\delta_1^V$ is the set of $\tau_1^V$-closed sets, is a basis for $\tau_2^V$.
%\end{proof}
\end{lem}
\begin{lem}\label{PH}
If $(S,\tau_1^S,\tau_2^S)$ is a pairwise Boolean space then $V_P(S)=(\mathcal{K}(S),\tau_1^V,\tau_2^V)$ is pairwise Hausdorff.
%\begin{proof}
%Let $C,C'\in\mathcal{K}(S)$ and $C\neq C'$. Let $z\in C$ such that $z\neq z'$, $\forall z'\in C'$. For each point $z'\in C'$, we choose disjoint open sets $U_{z'}^c\in\beta_2^S$ and $U_{z'}\in\beta_1^S$ (using the
%condition that $(S,\tau_1^S,\tau_2^S)$ is pairwise Hausdorff space.) containing points $z'$ and $z$, respectively. So the collection $\{U_{z'}^c:z'\in C'\}$ is $\tau_2^S$-open covering of $C'$. As $C'$ is pairwise compact, so there is a finite collection $\{U_{z'_{i}}^c:i=1,2,\cdots,n\}$ such that $C'\subseteq\bigcup_{i=1}^nU_{z'_{i}}^c$. Let $V'=\displaystyle\bigcup_{i=1}^nU_{z'_{i}}^c$ and $U=\displaystyle\bigcap_{i=1}^nU_{z'_{i}}$. As $z\in C\cap U$, hence $C\cap U\neq\emptyset$. Also, $C'\cap U=\emptyset$ because $C'\subseteq U^c$. It follows that $C\in\Diamond U\in\tau_1^V$ and $C'\notin\Diamond U$ i.e., $C'\in (\Diamond U)^c=\Box U^c\in\tau_2^V$. So we have two disjoint open sets $\Diamond U\in\tau_1^V$ and $\Box U^c\in\tau_2^V$ containing $C$ and $C'$, respectively. 
%\end{proof}
\end{lem}
\begin{lem}
\label{PC}
If $(S,\tau_1^S,\tau_2^S)$ is a pairwise Boolean space then $V_P(S)=(\mathcal{K}(S),\tau_1^V,\tau_2^V)$ is pairwise compact.
%\begin{proof}
%It is known from Proposition \ref{PROP1} that $\{\Box U,\Diamond U:U\in\beta_1^S\cup\beta_2^S\}$ is a subbasis for the topology $\tau_1^S\vee\tau_2^S$. We shall show that every cover of $\mathcal{K}(S)$ by subbasis-open sets has a finite subcover. Let $\mathcal{K}(S)=\bigcup_{\lambda\in\Lambda}\Box U_{\lambda}\cup\bigcup_{i\in I}\Diamond V_i$. Consider $S_1=S-\bigcup_{i\in I}V_i$. Then $S_1$ is a pairwise closed subset of $S$. Hence, $S_1\in\mathcal{K}(S)$. Since, $S_1\notin\Diamond V_i$ for each $i\in I$, so that $S_1\in\bigcup_{\lambda\in \Lambda}\Box U_{\lambda}$. Then for some $\lambda'\in\Lambda$, $S_1\in\Box U_{{\lambda}'}$. As a result, $S_1\subseteq U_{{\lambda}'}$ and hence $S-U_{{\lambda}'}\subseteq S-S_1=\displaystyle\bigcup_{i\in I}V_i$. Then, $S=U_{{\lambda}'}\cup\displaystyle\bigcup_{i\in I}V_i$. As $S$ is pairwise compact, we have $S=U_{\lambda'}\cup\bigcup_{i=1}^nV_i$. Let $A$ be an arbitrary element of $\mathcal{K}(S)$. If $A\subseteq U_{{\lambda}'}$ then $A\in\Box U_{{\lambda}'}$ otherwise $A\subseteq \bigcup_{i\in I}V_i$ i.e., $A\cap V_i\neq\emptyset$ for some $i\in \{1,2,\cdots, n\}$. As a result, $A\in \Box U_{\lambda'}\cup\bigcup_{i\in I}\Diamond V_i$. Therefore, $V_P(S)=(\mathcal{K}(S),\tau_1^V,\tau_2^V)$ is pairwise compact.
%\end{proof}
\end{lem}
Lemmas \ref{PZD}, \ref{PH} and \ref{PC} establish the following result:
\begin{thm}\cite{lauridsen2015bitopological}
\label{VPBS}
If $(S,\tau_1^S,\tau_2^S)$ is a pairwise Boolean space then $V_P(S)=(\mathcal{K}(S),\tau_1^V,\tau_2^V)$ is also a pairwise Boolean space. 
\end{thm}
We now extend the construction of bi-Vietoris functor on pairwise Stone spaces (cf. \cite{lauridsen2015bitopological}) to our $\mathcal{L}$-valued pairwise Boolean spaces setting by integrating the subalgebra-indexed, meet-preserving structure map $\alpha$ through $V_P$. Consequently, we obtain the $\mathcal{L}$-biVietoris functor $V_{\mathcal{L}}^{bi}$ on $PBS_{\mathcal{L}}$ and use it to prove the coalgebraic duality for the algebras of Fitting's $\mathcal{L}$-valued modal logic.
%We extend the construction of bi-Vietoris functor on pairwise Stone spaces 
%We now construct the $\mathcal{L}$-biVietoris functor $V_{\mathcal{L}}^{bi}$.
\begin{defn}
We define a $\mathcal{L}$-biVietoris functor $V_{\mathcal{L}}^{bi}:PBS_{\mathcal{L}}\to PBS_{\mathcal{L}}$ as follows:
\begin{enumerate}[(i)]
	\item For an object $(S,\tau_1^S,\tau_2^S,\alpha_S)$ in $PBS_{\mathcal{L}}$, we define $V_{\mathcal{L}}^{bi}(S,\tau_1^S,\tau_2^S,\alpha_S)=(V_P(S),V_P\circ\alpha_S)$ where $\alpha_S$ is a mapping from $\mathfrak{S}_{\mathcal{L}}$ to $\Lambda_S$, then for any $\mathcal{L}_1\in\mathfrak{S}_{\mathcal{L}}$, $V_P\circ\alpha_S(\mathcal{L}_1)$ is the pairwise Vietoris space of a pairwise closed subspace (i.e., pairwise Boolean subspace) $\alpha_S(\mathcal{L}_1)$ of $S$;
	\item For an arrow $f:(S_1,\tau_1^{S_1},\tau_2^{S_1},\alpha_{S_1})\to (S_2,\tau_1^{S_2},\tau_2^{S_2},\alpha_{S_2})$ in $PBS_{\mathcal{L}}$, $V_{\mathcal{L}}^{bi}(f):(V_P(S_1),V_P\circ\alpha_{S_1})\to (V_P(S_2),V_P\circ\alpha_{S_2})$ is defined by $V_{\mathcal{L}}^{bi}(f)(K)=f[K]$, where $K\in V_P(S_1)$.
\end{enumerate}
\end{defn}
\begin{rem}
We note that for $\mathcal{L}=\{0,1\}$, the two-element Boolean algebra, our $\mathcal{L}$-biVietoris structure recovers Lauridsen's bi-Vietoris construction, and if both the topologies $\tau_1$ and $\tau_2$ are equal then $\mathcal{L}$-biVietoris functor becomes classical Vietoris functor on the category of Stone spaces.
\end{rem}
We now show the well-definedness of the functor $V_{\mathcal{L}}^{bi}$.
\begin{lem}
\label{WDVLBI1}
Let $(S,\tau_1^S,\tau_2^S,\alpha_S)$ be an object in $PBS_{\mathcal{L}}$. Then $V_{\mathcal{L}}^{bi}(S,\tau_1^S,\tau_2^S,\alpha_S)$ is an object in $PBS_{\mathcal{L}}$.
\begin{proof}
Theorem \ref{VPBS} shows that $V_P(S)$ is a pairwise Boolean space. Now we shall show that $V_P\circ\alpha_S$ is a pairwise closed subspace of $V_P(S)$. For $\mathcal{L}_1\in\mathfrak{S}_{\mathcal{L}}$, an element of $V_P(S)\circ\alpha_S(\mathcal{L}_1)$ is a pairwise compact subset of $\alpha_S(\mathcal{L}_1)$. As $\alpha_S(\mathcal{L}_1)$ is also pairwise compact subspace of $S$, so that an element of $V_P\circ\alpha_S(\mathcal{L}_1)$ is a pairwise compact subset of $S$. As a result, $V_P\circ\alpha_S(\mathcal{L}_1)$ is a subset of $V_P(S)$. For $U\in\beta_1^S$, we get $\Box U\cap V_P\circ\alpha_S(\mathcal{L}_1)=\{C\in V_P\circ\alpha_S(\mathcal{L}_1):C\subset U\}=\Box (U\cap\alpha_S(\mathcal{L}_1))$ and $\Diamond U\cap V_P\circ\alpha_S(\mathcal{L}_1)=\{C\in V_P\circ\alpha_S(\mathcal{L}_1):C\cap U\neq\emptyset\}=\Diamond (U\cap\alpha_S(\mathcal{L}_1))$. Similarly for 
 $U\in\beta_2^S$. Hence, $V_P\circ\alpha_S(\mathcal{L}_1)$ is a pairwise subspace of $V_P(S)$. Since $\alpha_S(\mathcal{L}_1)$ is a pairwise Boolean subspace of $S$, by Theorem \ref{VPBS} we have $V_P\circ\alpha_S(\mathcal{L}_1)$ is a pairwise Boolean space. Henceforth, $V_P\circ\alpha_S(\mathcal{L}_1)$ is a pairwise closed subspace of $V_P(S)$.\\
Now we show that $V_P\circ\alpha_S$ satisfies the conditions given in the object part of Definition \ref{PBSL}. If $\alpha_S(\mathcal{L})=S$ then $V_P\circ\alpha_S(\mathcal{L})=V_P(S)$.\\
Let $\mathcal{L}_1,\mathcal{L}_2,\mathcal{L}_3\in\mathfrak{S}_{\mathcal{L}}$. If $\mathcal{L}_1=\mathcal{L}_2\cap \mathcal{L}_3$ then we show that $V_P(\alpha_S(\mathcal{L}_1))=V_P(\alpha_S(\mathcal{L}_2))\cap V_P(\alpha_S(\mathcal{L}_3))$. Now $V_P(\alpha_S(\mathcal{L}_1))=V_P(\alpha_S(\mathcal{L}_2\cap\mathcal{L}_3))=V_P(\alpha_S(\mathcal{L}_2)\cap\alpha_S(\mathcal{L}_3))$. The element structure of $V_P(\alpha_S(\mathcal{L}_2)\cap\alpha_S(\mathcal{L}_3))$ is of the form $P\cap (\alpha_S(\mathcal{L}_2)\cap\alpha_S(\mathcal{L}_3))$ and $Q\cap (\alpha_S(\mathcal{L}_2)\cap\alpha_S(\mathcal{L}_3))$, where $P$ and $Q$ are $\tau_1^S$-closed set and $\tau_2^S$-closed set, respectively. The elements of $V_P(\alpha_S(\mathcal{L}_2))\cap V_P(\alpha_S(\mathcal{L}_3))$ are of the form $(P_1\cap\alpha_S(\mathcal{L}_2))\cap (P_2\cap\alpha_S(\mathcal{L}_3))$ and $(Q_1\cap\alpha_S(\mathcal{L}_2))\cap (Q_2\cap\alpha_S(\mathcal{L}_3))$, where $P_1, P_2$ are $\tau_1^S$-closed and $Q_1, Q_2$ are $\tau_2^S$-closed. Then it is easy to show that $V_P(\alpha_S(\mathcal{L}_2)\cap\alpha_S(\mathcal{L}_3))\subseteq V_P(\alpha_S(\mathcal{L}_2))\cap V_P(\alpha_S(\mathcal{L}_3))$ and $V_P(\alpha_S(\mathcal{L}_2))\cap V_P(\alpha_S(\mathcal{L}_3))\subseteq V_P(\alpha_S(\mathcal{L}_2)\cap\alpha_S(\mathcal{L}_3))$.
\end{proof}
\end{lem}
\begin{lem}
\label{WDVLBI2}
Let $\eta:(S_1,\tau_1^{S_1},\tau_2^{S_1},\alpha_{S_1})\to (S_2,\tau_1^{S_2},\tau_2^{S_2},\alpha_{S_2})$ be an arrow in $PBS_{\mathcal{L}}$. Then $V_{\mathcal{L}}^{bi}(\eta)$ is an arrow in $PBS_{\mathcal{L}}$.
\begin{proof}
Given that $\eta$ is a pairwise continuous map from a pairwise Boolean space $S_1$ to a pairwise Boolean space $S_2$. Let $K\in V_P(S_1)$. Then $K$ is a pairwise closed subset of $S_1$ and hence $K$ is pairwise compact. Now $V_{\mathcal{L}}^{bi}(\eta)(K)=\eta[K]$ is a pairwise compact
subset of $S_2$. Since $S_2$ is a pairwise Boolean space, $\eta[K]$ is a pairwise closed subset of $S_2$. As a result, $V_{\mathcal{L}}^{bi}(\eta)(K)\in V_P(S_2)$. To show that $V_{\mathcal{L}}^{bi}(\eta)$ is pairwise continuous, let $C\in\beta_1^{S_2}$ and $D\in\beta_2^{S_2}$. Then 
\begin{equation*}
\begin{split}
V_{\mathcal{L}}^{bi}(\eta)^{-1}(\Box C)&=\{K\in V_P(S_1):V_{\mathcal{L}}^{bi}(\eta)(K)\in \Box C\}\\
&=\{K\in\mathcal{K}(S_1):\eta[K]\subseteq C\}\\
&=\{K\in\mathcal{K}(S_1):K\subseteq \eta^{-1}(C)\}\\
&=\Box \eta^{-1}(C)
\end{split}
\end{equation*}
and 
\begin{equation*}
\begin{split}
V_{\mathcal{L}}^{bi}(\eta)^{-1}(\Diamond C)&=\{K\in V_P(S_1):V_{\mathcal{L}}^{bi}(\eta)(K)\in \Diamond C\}\\
&=\{K\in\mathcal{K}(S_1):\eta[K]\cap C\neq\emptyset\}\\
&=\{K\in\mathcal{K}(S_1):K\cap \eta^{-1}(C)\neq\emptyset\}\\
&=\Diamond \eta^{-1}(C)
\end{split}
\end{equation*}
Similarly, $V_{\mathcal{L}}^{bi}(\eta)^{-1}(\Box D)=\Box \eta^{-1}(D)$ and $V_{\mathcal{L}}^{bi}(\eta)^{-1}(\Diamond D)=\Diamond \eta^{-1}(D)$. Therefore, $V_{\mathcal{L}}^{bi}(\eta)$ is pairwise continuous. It is still necessary to demonstrate that  $V_{\mathcal{L}}^{bi}(\eta)$ is subspace preserving. Let $M\in V_P\circ\alpha_{S_1}(\mathcal{L}_1)$, $\mathcal{L}_1\in\mathfrak{S}_{\mathcal{L}}$. Then $M\subseteq \alpha_{S_1}(\mathcal{L}_1)$. As $\eta$ is an arrow in $PBS_{\mathcal{L}}$, hence $\eta$ is a subspace preserving map. Thus, $\eta(M)\subseteq \alpha_{S_2}(\mathcal{L}_1)$. It shows that $V_{\mathcal{L}}^{bi}(\eta)(M)\subseteq\alpha_{S_2}(\mathcal{L}_1)$. Thus we have $V_{\mathcal{L}}^{bi}(\eta)(M)\in V_P\circ\alpha_{S_2}(\mathcal{L}_1)$.
\end{proof}
\end{lem}
We introduce two functors $\mathfrak{B}$ and $\mathfrak{C}$ between the categories $PRBS_{\mathcal{L}}$ and $COALG(V_{\mathcal{L}}^{bi})$ to show that these two categories are isomorphic.
\begin{defn}
\label{}
We define a functor $\mathfrak{B}:PRBS_{\mathcal{L}}\to COALG(V_{\mathcal{L}}^{bi})$ as follows:
\begin{enumerate}[(i)]
\item For an object $(S,\alpha_S,\mathcal{R})$ in $PRBS_{\mathcal{L}}$, define $\mathfrak{B}(S,\alpha_S,\mathcal{R})=(S,\alpha_S,\mathcal{R}[-])$, where $\mathcal{R}[-]:(S,\alpha_S)\to V_{\mathcal{L}}^{bi}(S,\alpha_S)$ is an arrow in $PBS_{\mathcal{L}}$ defined by $\mathcal{R}[s]=\{p\in S:s\mathcal{R}p\}$, $s\in S$;
\item For an arrow $f:(S_1,\alpha_{S_1},\mathcal{R}_1)\to (S_2,\alpha_{S_2},\mathcal{R}_2)$ in $PRBS_{\mathcal{L}}$, define $\mathfrak{B}(f):(S_1,\alpha_{S_1},\mathcal{R}_1[-])\to (S_2,\alpha_{S_2},\mathcal{R}_2[-])$ by $\mathfrak{B}(f)=f$.
\end{enumerate}
\end{defn}
The well-definedness of the functor $\mathfrak{B}$ is shown by the following two lemmas:
\begin{lem}
\label{WDB1}
Let $(S,\alpha_S,\mathcal{R})$ be an object in $PRBS_{\mathcal{L}}$. Then $\mathfrak{B}(S,\alpha_S,\mathcal{R})$ is an object in $COALG(V_{\mathcal{L}}^{bi})$ .
\begin{proof}
We shall show that $\mathcal{R}[-]:(S,\alpha_S)\to V_{\mathcal{L}}^{bi}(S,\alpha_S)$ is an arrow in $PBS_{\mathcal{L}}$. By the conditions given in the object part of Definition \ref{PRBSL}, we know that for each $s\in S$, $\mathcal{R}[s]$ is pairwise compact subset of $S$. As $S$ is pairwise Boolean space, hence $\mathcal{R}[s]$ is a pairwise closed subset of $S$. Thus $\mathcal{R}[s]\in V_P(S)$. Let $U\in\beta_1^S$. Then 
\begin{equation*}
\begin{split}
\mathcal{R}[-]^{-1}(\Box U) &=\{s\in S:\mathcal{R}[s]\in\Box U\}\\
&=\{s\in S:\mathcal{R}[s]\subseteq U\}\\
&=[\mathcal{R}]U\in\beta_1^S\text{ [by Definition \ref{PRBSL}]}
\end{split}
\end{equation*}
and 
\begin{equation*}
\begin{split}
\mathcal{R}[-]^{-1}(\Diamond U) &=\{s\in S:\mathcal{R}[s]\in\Diamond U\}\\
&=\{s\in S:\mathcal{R}[s]\cap U\neq\emptyset\}\\
&=\langle\mathcal{R}\rangle U\in\beta_1^S \text{ [by Definition \ref{PRBSL}]}
\end{split}
\end{equation*}
Similarly, for $U\in\beta_2^S$, $\mathcal{R}[-]^{-1}(\Box U)=[\mathcal{R}]U\in\beta_2^S$ and $\mathcal{R}[-]^{-1}(\Diamond U)=\langle\mathcal{R}\rangle U\in\beta_2^S$. Henceforth, $\mathcal{R}[-]$ is a pairwise continuous map. Now we show that $\mathcal{R}[-]$ is subspace preserving. Let $s\in \alpha_S(\mathcal{L}')$, $\mathcal{L}'\in\mathfrak{S}_{\mathcal{L}}$. It is known from Definition \ref{PRBSL} that $\mathcal{R}[s]$ is a 
pairwise compact subset of $\alpha_S(\mathcal{L}')$. Since $\alpha_S(\mathcal{L}')$ is itself a pairwise Boolean space, thus we have $\mathcal{R}[s]\in V_P\circ \alpha_S(\mathcal{L}')$. Therefore, $\mathfrak{B}(S,\alpha_S,\mathcal{R})$ is a $V_{\mathcal{L}}^{bi}$-coalgebra.
\end{proof}
\end{lem}
\begin{lem}
\label{WDB2}
Let $f:(S_1,\alpha_{S_1},\mathcal{R}_1)\to (S_2,\alpha_{S_2},\mathcal{R}_2)$ be an arrow in $PRBS_{\mathcal{L}}$. Then $\mathfrak{B}(f)$ is an arrow in $COALG(V_{\mathcal{L}}^{bi})$.
\begin{proof}
As $f$ is an arrow in $PRBS_{\mathcal{L}}$, $\mathfrak{B}(f)=f:(S_1,\alpha_{S_1},\mathcal{R}_1[-])\to (S_2,\alpha_{S_2},\mathcal{R}_2[-])$ is a  pairwise continuous map. Now using the conditions mentioned in the arrow part
of Definition \ref{PRBSL}, it is straightforward to verify that $\mathcal{R}_2[-]\circ f=V_{\mathcal{L}}^{bi}\circ\mathcal{R}_1[-]$. Thus $\mathfrak{B}(f)$ is an arrow in $COALG(V_{\mathcal{L}}^{bi})$.
\end{proof}
\end{lem}
\begin{defn}
\label{}
We define a functor $\mathfrak{C}:COALG(V_{\mathcal{L}}^{bi})\to PRBS_{\mathcal{L}}$ as follows:
\begin{enumerate}[(i)]
\item For an object $((C,\alpha_C),\xi)$ in $COALG(V_{\mathcal{L}}^{bi})$, define $\mathfrak{C}((C,\alpha_C),\xi)=(C,\alpha_C,\mathcal{R}_{\xi})$, where $\mathcal{R}_{\xi}$ is a binary relation on $C$ defined by $d\in\mathcal{R}_{\xi}[c]\iff d\in\xi(c)$, $c,d\in C$;
\item For an arrow $f:((C_1,\alpha_{C_1}),\xi_1)\to ((C_2,\alpha_{C_2}),\xi_2)$ in $COALG(V_{\mathcal{L}}^{bi})$, define $\mathfrak{C}(f):(C_1,\alpha_{C_1},\mathcal{R}_{\xi_1})\to (C_2,\alpha_{C_2},\mathcal{R}_{\xi_2})$ by $\mathfrak{C}(f)=f$.
\end{enumerate}
\end{defn}
The well-definedness of the functor $\mathfrak{C}$ is shown by Lemma \ref{WDC1} and Lemma 
\ref{WDC2}.
\begin{lem}
\label{WDC1}
For an object $((C,\alpha_C),\xi)$ in $COALG(V_{\mathcal{L}}^{bi})$, $\mathfrak{C}((C,\alpha_C),\xi)=(C,\alpha_C,\mathcal{R}_{\xi})$ is an object in $PRBS_{\mathcal{L}}$.
\begin{proof}
In order to show that $\mathfrak{C}((C,\alpha_C),\xi)$ is an object in $PRBS_{\mathcal{L}}$, we must verify that $\mathfrak{C}((C,\alpha_C),\xi)$ satisfies the conditions given in the object part of Definition \ref{PRBSL}. For each $c\in C$, $\mathcal{R}_{\xi}[c]=\xi(c)\in V_P(C)$. Hence, $\mathcal{R}_{\xi}[c]$ is a pairwise closed subset of $C$. Thus $\mathcal{R}_{\xi}[c]$ is pairwise compact. Let $D\in\beta_1^C$. Then 
\begin{equation*}
\begin{split}
[\mathcal{R}_{\xi}](D)& =\{c\in C:\mathcal{R}_{\xi}[c]\subseteq D\}\\
&=\{c\in C:\xi(c)\subseteq D\}\\
&=\{c\in C:\xi(c)\in\Box D\}\\
&=\xi^{-1}(\Box D)\in\beta_1^C
\end{split}
\end{equation*}
and 
\begin{equation*}
\begin{split}
\langle \mathcal{R}_{\xi}\rangle D &=\{c\in C:\mathcal{R}_{\xi}[c]\cap D\neq \emptyset\}\\
&=\{c\in C:\xi(c)\cap D\neq\emptyset\}\\
&=\{c\in C:\xi(c)\in\Diamond D\}\\
&=\xi^{-1}(\Diamond D)\in\beta_1^C
\end{split}
\end{equation*}
Finally, let $m\in\alpha_C(\mathcal{L'})$ for $\mathcal{L'}\in\mathfrak{S}_{\mathcal{L}}$. As $\xi$ is a subspace preserving map from $(C,\alpha_C)$ to $V_{\mathcal{L}}^{bi}(C,\alpha_C)$, we have $\mathcal{R}_{\xi}[m]=\xi(m)\in V_P\circ\alpha_C(\mathcal{L'})$. Henceforth, $\mathcal{R}_{\xi}[m]\subset\alpha_C(\mathcal{L'})$.
\end{proof}
\end{lem}
\begin{lem}
\label{WDC2}
For an arrow $f:((C_1,\alpha_{C_1}),\xi_1)\to((C_2,\alpha_{C_2}),\xi_2)$ in $COALG(V_{\mathcal{L}}^{bi})$, $\mathfrak{C}(f):(C_1,\alpha_{C_1},\mathcal{R}_{\xi_1})\to (C_2,\alpha_{C_2},\mathcal{R}_{\xi_2})$ is an arrow in $PRBS_{\mathcal{L}}$.
\begin{proof}
It is straightforward to prove that $\mathfrak{C}$ is an arrow in $PRBS_{\mathcal{L}}$.
\end{proof}
\end{lem}
Now we obtain the following result:
\begin{thm}
\label{PCI}
The categories $PRBS_{\mathcal{L}}$ and $COALG(V_{\mathcal{L}}^{bi})$ are isomorphic.
\begin{proof}
We shall show that the categories $PRBS_{\mathcal{L}}$ and $COALG(V_{\mathcal{L}}^{bi})$ are isomorphic via the functors $\mathfrak{B}$ and $\mathfrak{C}$. Let $(S,\alpha_S,\mathcal{R})$ be an object in $PRBS_{\mathcal{L}}$. Then $\mathfrak{C}\circ\mathfrak{B}(S,\alpha_S,\mathcal{R})=\mathfrak{C}(S,\alpha_S,\mathcal{R}[-])=(S,\alpha_S,\mathcal{R}_{\mathcal{R}[-]})$. Now $t\in \mathcal{R}_{\mathcal{R}[-]}(s)\iff t\in \mathcal{R}[s]$. Thus, $(S,\alpha_S,\mathcal{R})=\mathfrak{C}\circ\mathfrak{B}(S,\alpha_S,\mathcal{R})$. Let $((C,\alpha_C),\xi)$ be an object in $COALG(V_{\mathcal{L}}^{bi})$. Then $\mathfrak{B}\circ\mathfrak{C}((C,\alpha_C),\xi)=\mathfrak{B}(C,\alpha_C,\mathcal{R}_{\xi})=((C,\alpha_C),\mathcal{R}_{\xi}[-])$. We have $c_2\in\mathcal{R}_{\xi}[c_1]\iff c_2\in\xi(c_1)$. As a result, $((C,\alpha_C),\xi)=\mathfrak{B}\circ\mathfrak{C}((C,\alpha_C),\xi)$. It is clear that for an arrow $f$ in $COALG(V_{\mathcal{L}}^{bi})$, $\mathfrak{B}\circ\mathfrak{C}(f)=f$ and for an arrow $f$ in $PRBS_{\mathcal{L}}$, $\mathfrak{C}\circ\mathfrak{B}(f)=f$.
\end{proof}
\end{thm}
Using Theorems \ref{DRML} and \ref{PCI}, we arrive at the following duality theorem: 
\begin{thm}
\label{DRMLC}
The categories $\mathcal{MA}_{\mathcal{L}}$ and $COALG(V_{\mathcal{L}}^{bi})$ are dually equivalent.
\end{thm}
Thus the modal semi-primal duality for algebras of Fitting’s Heyting-valued 
modal logic (for more information, see \cite{maruyama2011dualities}) can potentially be represented in terms of the coalgebras of $\mathcal{L}$-biVietoris functor $V_{\mathcal{L}}^{bi}$.\\
For $\mathcal{L}=\mathbf{2}$, the two-element Boolean algebra, our duality which is established in Section \ref{BDFML} gives the J{\'o}nsson-Tarski duality for modal algebras. When composed with Theorem \ref{DRMLC}, this provides $\mathcal{MA}_{\mathbf{2}}\simeq COALG(V_{\mathbf{2}}^{bi})^{op}$, which recovers the Abramsky-Kupke-Kurz-Venema coalgebraic duality.\\
Finally, based on the preceding theorems (Theorems \ref{DRML},\ref{PCI},\ref{DRMLC}), we can conclude:
\begin{thm}
	Fitting's multi-valued modal logic is sound and complete in the sense of coalgebras of the $\mathcal{L}$-biVietoris functor $V_{\mathcal{L}}^{bi}$.
\end{thm}

	The same methodology can be used to build both bitopological and coalgebraic duality for {\L}ukasiewicz $n$-valued modal logic, where $n=\{0,\frac{1}{n-1},\frac{2}{n-2},\cdots,\frac{n-2}{n-1}, 1 \}$. It should be noted that  $(n, \vee,\wedge,\star,\circledast,\rightarrow,()^c, 0, 1)$ is a semi-primal algebra, where the operations $( \vee,\wedge,\star,\circledast,\rightarrow,()^c, 0, 1)$ are defined by 
	\begin{equation*}
		\begin{aligned}
		p\vee q&=max(p,q)\\
		p\wedge q&=min(p,q)\\
		p\star q& =max(0,p+q-1)\\
		p\circledast q&=min(1,p+q)\\
		p\rightarrow q&=min(1,1-p+q)\\
		p^c& =1-p
		\end{aligned}
	\end{equation*}
and $0,1$ are considered as constants, i.e., nullary-operations. Algebras of {\L}ukasiewicz-$n$-valued modal logic can be found in \cite{maruyama2012natural}.\\

Thus, it is easy to follow, with sort of adjustment, that bitopological and coalgebraic duality can be developed for the general algebraic structure $\mathbb{ISP_M}(\mathcal{L})$, $\mathcal{L}$ is a finite algebra, using the same method as explained in this paper.
\section{Applications}\label{LAB}
In this section we provide fundamental applications of our coalgebraic approach to the bitopological duality.\\
%As an application of this coalgebraic approach to the bitopological duality, we establish the following:
\textbf{Hennessy-Milner Theorem for Fitting's $\mathcal{L}$-valued modal logic }
The classical Hennessy-Milner theorem \cite{hennessy1985algebraic} states that modal equivalence coincides with bisimulation for image finite Kripke models. In this study, we establish Hennessy-Milner theorem for Fitting's $\mathcal{L}$-valued modal logic based on our bitopological duality $\mathcal{MA}_{\mathcal{L}}\simeq PRBS_{\mathcal{L}}^{op}$ and $PRBS_{\mathcal{L}}\cong COALG(V_{\mathcal{L}}^{bi})$, and using the canonical coalgebra on the dual space obtained by J{\'o}nsson-Tarski duality. By our duality $\mathcal{MA}_{\mathcal{L}}\simeq PRBS_{\mathcal{L}}^{op}$, the dual of a $\mathcal{L}$-$\mathcal{ML}$-algebra $A$ is the canonical object $\mathcal{G}(A)=(HOM_{\mathcal{VA}_{\mathcal{L}}}(A,\mathcal{L}),\tau_1,\tau_2,\alpha_A,\mathcal{R}_{\Box})$, where $\mathcal{R}_{\Box}$ is uniquely determined by the J{\'o}nsson-Tarski transport of $\Box$-modality: $[\mathcal{R}_{\Box}]\langle U_\ell(a)\rangle=\langle U_\ell(\Box a)\rangle$, $\forall a\in A$ and $\ell\in \mathcal{L}$, $\langle U_\ell (a)\rangle=\{h\in HOM_{\mathcal{VA}_{\mathcal{L}}}(A,\mathcal{L}):h(a)\geq\ell\}$. The categorical equivalence $PRBS_{\mathcal{L}}\simeq COALG(V_{\mathcal{L}}^{bi})$ yields the $V_{\mathcal{L}}^{bi}$-coalgebra structure $(\mathcal{G}(A),\zeta)$ such that $[\zeta](\langle U_\ell(a)\rangle)=\langle U_\ell(\Box a)\rangle$. By defining the theory map into this $V_{\mathcal{L}}^{bi}$-coalgebra, we develop Hennessy-Milner theorem for $V_{\mathcal{L}}^{bi}$-coalgebra-models without image-finiteness concept. As a result, it generalizes the classical Hennessy-Milner theorem for Vietoris coalgebra on Stone spaces \cite{bezhanishvili2010vietoris}.

\textbf{Language and models}\\
We fix Fitting's $\mathcal{L}$-valued modal logic $\mathcal{L}$-$\mathcal{ML}$ (cf. \cite{maruyama2009algebraic}). For the purposes of this section, we review its language and semantics. \\

The signature of $\mathcal{L}$-$\mathcal{ML}$ consists of a unary modal operation $\Box$, and the unary operation $T_a (a\in \mathcal{L})$, the binary operations $\wedge, \vee, \rightarrow$, the nullary constants $\bot (0),\top(1)$. Let $\mathsf{Prop}$ denote the set of propositional variables. The set $\mathrm{Form}_{\Box}$ of formulas of $\mathcal{L}$-$\mathcal{ML}$ is generated by:
\[
\varphi ::= p \mid \top\mid\bot \mid (\varphi\wedge\psi)\mid(\varphi\vee\psi)\mid (\varphi\rightarrow \psi)\mid T_\ell(\varphi)\ (\ell\in \mathcal{L})\mid \Box\varphi,
\qquad p\in\mathsf{Prop}.
\]

\begin{defn}[$PRBS_{\mathcal{L}}$-model]
 	A $PRBS_{\mathcal{L}}$-model is defined by $M=(B,\alpha_B,\mathcal{R},v)$, where 
 \begin{enumerate}[(i)]
 	\item $(B,\alpha_B)$ is an object in $PBS_{\mathcal{L}}$;
 	\item $\mathcal{R}$ is the binary relation on $B$ that is compatible with both the topologies defined on $B$ and the structure map $\alpha_B$;
 	\item $v:\mathsf{Prop}\to \mathcal{L}^B$ is a $\mathcal{L}$-valuation map such that for each $p\in \mathsf{Prop}$ and every $\ell\in\mathcal{L}$, the set of states where the value is at least $\ell$, i.e., the set $\{x\in B:v(p)(x)\geq \ell\}$, is pairwise closed in $B$ and for every subalgebra $L'$ of $\mathcal{L}$, $v(p)[\alpha_B(L')]\subseteq L'$.
 	 \end{enumerate}
\end{defn}
As the category $PRBS_{\mathcal{L}}$ isomorphic to $COALG(V_{\mathcal{L}}^{bi})$, $PRBS_{\mathcal{L}}\cong COALG(V_{\mathcal{L}}^{bi})$, a coalgebra $(B,\alpha_B,\xi_B)$ corresponds to a $\mathcal{L}$-valued relational space $(B,\alpha_B,\mathcal{R})$.
\begin{defn}
	Let $(\mathbb{B},\xi_{\mathbb{B}},v_{\mathbb{B}})$ and $(\mathbb{D},\xi_{\mathbb{D}},v_{\mathbb{D}})$ be two $V_{\mathcal{L}}^{bi}$-coalgebra-models. A mapping $f:\mathbb{B}\to\mathbb{D}$ is a $V_{\mathcal{L}}^{bi}$-coalgebra-model morphism if 
	\begin{enumerate}[(i)]
		\item $f:\mathbb{B}\to\mathbb{D}$ is an arrow in $PBS_{\mathcal{L}}$;
		\item$f$ is a $V_{\mathcal{L}}^{bi}$-coalgebra morphism ;
		\item $f$ preserves atomic valuations: $\forall p\in\mathsf{Prop}$, $v_{\mathbb{D}}(p)\circ f=v_{\mathbb{B}}(p)$.
	\end{enumerate}
	The collection of $V_{\mathcal{L}}^{bi}$-coalgebra-models and $V_{\mathcal{L}}^{bi}$-coalgebra-model morphisms forms a category, denoted by $\mathsf{MOD}(V_{\mathcal{L}}^{bi})$.
\end{defn}
\textbf{Interpretation of formulas}\\
Given a model $M=(B,\alpha_B,\mathcal{R},v)$, the interpretation map (evaluation map) $\llbracket\cdot\rrbracket_M:\mathrm{Form}_{\Box}\to\mathcal{L}^B$ is the unique map that extends $\mathcal{L}$-valuation map $v$ on $\mathsf{Prop}$ and is determined by the following clauses for every state $x\in B$: 
%each formula $\varphi\in \mathrm{Form}_{\Box}$ is represented by a map $\llbracket\varphi\rrbracket_M:B\to\mathcal{L}$
%We extend the $\mathcal{L}$-valuation map $v$ to all formulas $\mathrm{Form}_{\Box}$ by an interpretation map $\llbracket \cdot \rrbracket: \mathrm{Form}_{\Box}\to \mathcal{L}^B$ as follows:
\begin{enumerate}[(i)]
	\item $\llbracket p\rrbracket_{M}=v(p)$ ;
	\item $\llbracket \bot\rrbracket_{M}(x)=0_{\mathcal{L}}$, $\llbracket \top\rrbracket_{M}(x)=1_{\mathcal{L}}$;
		\item $\llbracket \varphi\wedge\psi\rrbracket_{M}(x)=\llbracket \varphi\rrbracket_{M}(x)\wedge \llbracket \psi\rrbracket_{M}(x)$;
		\item $\llbracket \varphi\vee\psi\rrbracket_{M}(x)=\llbracket \varphi\rrbracket_{M}(x)\vee \llbracket \psi\rrbracket_{M}(x)$;
		\item $\llbracket\varphi\rightarrow\psi\rrbracket_M(x)=\llbracket\varphi\rrbracket_M(x)\Rightarrow_{\mathcal{L}}\llbracket\psi\rrbracket_{M}(x)$;
		\item $\llbracket T_\ell(\varphi)\rrbracket_{M}(x)=T_\ell(\llbracket\varphi\rrbracket_{M}(x))$;
		\item $\llbracket\Box\varphi\rrbracket_{M}(x)=\bigwedge\{\llbracket\varphi\rrbracket_{M}(y):x\mathcal{R} y\}$.
		\end{enumerate}
\begin{note}
The set of formulas $\mathrm{Form}_{\Box}$ is the free algebra over $\mathsf{Prop}$ in the variety of $\mathcal{L}$-$\mathcal{ML}$-algebras. The set $\mathcal{L}^B$ carries the structure of a $\mathcal{L}$-$\mathcal{ML}$-algebra with the connectives ($\vee,\wedge,\rightarrow,T_a (a\in\mathcal{L}),0,1$) defined pointwise and the modal connective $\Box$ defined by $(\Box f)(x)=\bigwedge\{f(y):y\in \mathcal{R}[x]\}$. The universal property of the free algebra gives that there is a unique homomorphism $\llbracket\cdot\rrbracket_M:\mathrm{Form}_{\Box}\to\mathcal{L}^B$ that extends the valuation $v$ on atomic propositions. Therefore, the interpretation map is unique.
\end{note}
\begin{rem}
	\begin{enumerate}[(i)]
		\item Our $PRBS_{\mathcal{L}}$-model is a topologically enriched $\mathcal{L}$-valued Kripke model of Maruyama \cite{maruyama2009algebraic}. The structure of the $PRBS_{\mathcal{L}}$-model ensures that for every formula $\varphi$ and truth value $a\in\mathcal{L}$, the designated region $\{x\in B:\llbracket\varphi\rrbracket_M(x)\geq a\}=\llbracket U_a(\varphi)\rrbracket_M$ is pairwise closed and for every sublagebra $K$ of $\mathcal{L}$, $\llbracket\varphi\rrbracket_M(\alpha_B(K))\subseteq K$.\\
		%If we remove both the topologies from the states set and the structure map $\alpha$, then any $PRBS_{\mathcal{L}}$-model becomes Maruyama's $\mathcal{L}$-valued Kripke model.
		Forgetting both the topologies and the structure map $\alpha_B$ from a $PRBS_{\mathcal{L}}$-model $(B,\alpha_B,\mathcal{R},v)$ yields the Maruyama's $\mathcal{L}$-valued Kripke model $(B,\mathcal{R},V)$ with the same truth definition: $V(x,p)=v(p)(x)$, $p\in\mathsf{Prop}$ and for any formulas $\varphi$, $V(x,\varphi)=\llbracket\varphi\rrbracket_{(B,\alpha_B,\mathcal{R},v)}(x)$.\\
		If the set of states $M$ is finite, the assignment $B=M$ with the two discrete Boolean topologies, the trivial structure map $\alpha_B$, i.e., $\alpha_B(K)=B, \forall K\in\mathfrak{S}_{|mathcal{L}}$, and $v(p)(x)=V(x,p)$ yields a $PRBS_{\mathcal{L}}$-model $(B,\tau,\tau,\alpha_B,\mathcal{R},v)$ whose induced interpretation map agrees with $V$ on all formulas.\\
		For an infinite $\mathcal{L}$-valued Kripke frame, we utilise the Stone duality framework to embed the model into a $PRBS_{\mathcal{L}}$-model. Consider an infinite $\mathcal{L}$-valued Kripke model $(\mathcal{M},\mathcal{R},V)$. Let $\mathbf{B}$ be the Boolean algebra generated by the sets $\{m\in\mathcal{M}:V(m,p)\geq a\}$, $p\in\mathsf{Prop}$ and $a\in\mathcal{L}$. Let $X=Ulf(\mathcal{B})$ ($Ulf(\mathcal{B})$ denotes the set of all ultrafliters of the Boolean algebra $\mathbf{B}$) be the Stone space of $\mathcal{B}$ and $\mathcal{E}: \mathcal{M}\to X$ be the canonical embedding (principal ultrafilter embeding). We endow $X$ with two Stone topologies: $\tau_1$ is the Stone topology of $\mathbf{B}$ with basis $\mathbb{B}_1=\{\mathcal{O}_a\mid a\in \mathbf{B}\}$, where $\mathcal{O}_a=\{U\in Ulf(\mathbf{B}):a\in U\}$, and $\tau_2$ is the topology with basis $\mathbb{B}_2=\{X\backslash \mathcal{O}_a:a\in \mathbf{B}\}$. Let $\widehat{A}=\{u\in X:A\in u\}$, where $A\in\mathbf{B}$. Then $\widehat{A}$ is clopen in both the Stone topologies on $X$.	We consider trivial structure map $\alpha_X:\mathfrak{S}_{\mathcal{L}}\to \Lambda_X$, i.e., $\alpha_X(L)=X$, for any subalgebra $L$ of $\mathcal{L}$. For each $p\in\mathsf{Prop}$, we define the valuation $v^*(p):X\to\mathcal{L}$ such that for each $a\in \mathcal{L}$, the set $\{x\in X: V^*(p)(x)\geq a\}=\widehat{ \{m\in\mathcal{M}:V(m,p)\geq a\}}$. Let $\mathcal{R}^*$ be a relation on $X$ such that for each $A\in\mathbf{B}$ $[\mathcal{R}^*](\widehat{A})=\widehat{[\mathcal{R}]A}$, $\langle\mathcal{R}^*\rangle\widehat{A}=\widehat{\langle\mathcal{R}\rangle A}$, where $[\mathcal{R}]A=\{m\in \mathcal{M}:\mathcal{R}[m]\subseteq A\}$ and $\langle\mathcal{R}\rangle A=\{m\in M:\mathcal{R}[m]\cap A\neq \emptyset\}$. Then $(X,\alpha_X,\mathcal{R}^*,v^*)$ is a $PRBS_{\mathcal{L}}$-model and the canonical embedding $\mathcal{E}:M\to X$ preserves truth values of all formulas, i.e.,  $\llbracket\varphi\rrbracket_{(X,\alpha_X,\mathcal{R}^*,v^*)}(\mathcal{E}(m))=V(m,\varphi)$, $\forall m\in\mathcal{M}, \varphi$. \\
		Thus, $PRBS_{\mathcal{L}}$-model is an extension of Maruyama's $\mathcal{L}$-valued Kripke model.
		\item When $\mathcal{L}=\{0,1\}$ (the two-element Boolean algebra) the only subalgebra of $\mathcal{L}$ is $\mathcal{L}$ itself and hence the structure map $\alpha$ is trivial. When $\mathcal{L}=\mathbf{2}$, we observe that on the canonical dual spaces obtained by applying the duality functor to the $\mathcal{L}$-$\mathcal{VL}$-algebras $\mathcal{A}$, both topologies coincide. Thus, objects in the category $PRBS_{\mathbf{2}}$ are precisely descriptive general frames. Hence, $PRBS_{\mathbf{2}}$-model is a descriptive general model.
	\end{enumerate}
\end{rem}
\begin{defn}[Modal equivalence]
	Let $\mathcal{M}=(B,\alpha_B,\mathcal{R},v)$ be a $PRBS_{\mathcal{L}}$-model. Any two states $x,y\in B$ are said to be modally equivalent, write $x\equiv y$ iff 
	\[
	\forall \varphi\in\mathrm{Form}_{\Box}\;\; \llbracket\varphi\rrbracket_{\mathcal{M}}(x)\;=\; \llbracket\varphi\rrbracket_{\mathcal{M}}(y).
	\]
\end{defn}

	The following lemma shows that $V_{\mathcal{L}}^{bi}$-coalgebra-model morphisms preserve truth.
	\begin{lem}\label{VPT}
		Let $h:(\mathbb{B},\xi_{\mathbb{B}},v_{\mathbb{B}})\to (\mathbb{D},\xi_{\mathbb{D}},v_{\mathbb{D}})$ be a $V_{\mathcal{L}}^{bi}$-coalgebra-model morphism. Then for every formula $\varphi\in\mathrm{Form}_\Box$ and every state $b\in\mathbb{B}$,
		\[
		\llbracket\varphi\rrbracket_{(\mathbb{B},\xi_{\mathbb{B}},v_{\mathbb{B}})}(b)=\llbracket\varphi\rrbracket_{(\mathbb{D},\xi_{\mathbb{D}},v_{\mathbb{D}})}(h(b)).
		\]
		Equivalently, for every $\ell\in\mathcal{L}$,
			\[
		\llbracket\Box\psi\rrbracket_{(\mathbb{B},\xi_{\mathbb{B}},v_{\mathbb{B}})}(b)\geq\ell=\llbracket\Box\psi\rrbracket_{(\mathbb{D},\xi_{\mathbb{D}},v_{\mathbb{D}})}(h(b))\geq\ell.
		\]
		\begin{proof}
			We prove it by induction on formulas. 
			\begin{enumerate}[(i)]
				\item $\varphi=p$,$p\in\mathsf{Prop}$. As $h$ is a $V_{\mathcal{L}}^{bi}$-coalgebra-model morphism, we have $v_{\mathbb{D}}(p)\circ h=v_{\mathbb{B}}(p)$. Hence, $\llbracket p\rrbracket_{(\mathbb{D},\xi_{\mathbb{D}},v_{\mathbb{D}})}(h(b))=\llbracket p\rrbracket_{(\mathbb{B},\xi_{\mathbb{B}},v_{\mathbb{B}})}(b)$.\\
				$\llbracket 0\rrbracket_{(\mathbb{B},\xi_{\mathbb{B}},v_{\mathbb{B}})}(b)=0=\llbracket 0\rrbracket_{(\mathbb{D},\xi_{\mathbb{D}},v_{\mathbb{D}})}(h(b))$ and $\llbracket 1\rrbracket_{(\mathbb{B},\xi_{\mathbb{B}},v_{\mathbb{B}})}(b)=1=\llbracket 1\rrbracket_{(\mathbb{D},\xi_{\mathbb{D}},v_{\mathbb{D}})}(h(b))$.
				\item Suppose the equality holds for formulas $\varphi, \psi$. then it is easy to follow that the equality holds for $\varphi\vee\psi,\varphi\wedge\psi,\varphi\rightarrow\psi$ and $T_\ell(\varphi)$, $\ell\in\mathcal{L}$.\\
				To prove the equality for a modal formula $\Box\psi$, we show that for any $b\in \mathbb{B}$ and $\ell\in\mathcal{L}$,
				\[
				\llbracket\Box\psi\rrbracket_{(\mathbb{B},\xi_{\mathbb{B}},v_{\mathbb{B}})}(b)\geq\ell=\llbracket\Box\psi\rrbracket_{(\mathbb{D},\xi_{\mathbb{D}},v_{\mathbb{D}})}(h(b))\geq\ell.
				\]
				Let $b\in\mathbb{B}$ and $\ell\in\mathcal{L}$. Now,
				$\llbracket\Box\psi\rrbracket_{(\mathbb{B},\xi_{\mathbb{B}},v_{\mathbb{B}})}\geq\ell=\displaystyle\bigwedge_{y\in\xi_{\mathbb{B}}(b)}\llbracket\psi\rrbracket_{(\mathbb{B},\xi_{\mathbb{B}},v_{\mathbb{B}})}(y)\geq\ell$. Thus 
				\[
				\llbracket\Box\psi\rrbracket_{(\mathbb{B},\xi_{\mathbb{B}},v_{\mathbb{B}})}\geq\ell\iff\xi_{\mathbb{B}}(b)\subseteq \mathfrak{S}_{\mathbb{B}},
				\]
				where $\mathfrak{S}_{\mathbb{B}}=\{y\in\mathbb{B}:\llbracket\psi\rrbracket_{(\mathbb{B},\xi_{\mathbb{B}},v_{\mathbb{B}})}(y)\geq\ell\}$. By the induction law, we have $\llbracket\psi\rrbracket_{(\mathbb{B},\xi_{\mathbb{B}},v_{\mathbb{B}})}=\llbracket\psi\rrbracket_{(\mathbb{D},\xi_{\mathbb{D}},v_{\mathbb{D}})}\circ h$. Consequently, $\mathfrak{S}_{\mathbb{B}}=h^{-1}(\mathfrak{S}_{\mathbb{D}})$, where $\mathfrak{S}_{\mathbb{D}}=\{z\in\mathbb{D}:\llbracket\psi\rrbracket_{(\mathbb{D},\xi_{\mathbb{D}},v_{\mathbb{D}})}(z)\geq\ell\}$. Since $h$ is a $V_{\mathcal{L}}^{bi}$-coalgebra morphism: $\xi_{\mathbb{D}}\circ h=V_{\mathcal{L}}^{bi}\circ\xi_{\mathbb{B}}$, hence, for all $b\in\mathbb{B}$, $\xi_{\mathbb{D}}(h(b))=V_{\mathcal{L}}^{bi}(h)(\xi_{\mathbb{B}}(b))=h[\xi_{\mathbb{B}}(b)]$.\\
				So,
				\[
				\begin{aligned}
				\llbracket\Box\psi\rrbracket_{(\mathbb{B},\xi_{\mathbb{B}},v_{\mathbb{B}})}(b)\geq\ell& \iff\xi_{\mathbb{B}}(b)\subseteq \mathfrak{S}_{\mathbb{B}}\\
				&\iff \xi_{\mathbb{D}}(h(b))\subseteq \mathfrak{S}_{\mathbb{D}}\\
				&\iff \xi_{\mathbb{D}}(h(b))\subseteq \mathfrak{S}_{\mathbb{D}}\\
				&\iff \llbracket\Box\psi\rrbracket_{(\mathbb{D},\xi_{\mathbb{D}},v_{\mathbb{D}})}(h(b))\geq \ell.
			\end{aligned}
				\]
				As this holds for every $\ell\in\mathcal{L}$, so $\llbracket\Box\psi\rrbracket_{(\mathbb{B},\xi_{\mathbb{B}},v_{\mathbb{B}})}(b)=\llbracket\Box\psi\rrbracket_{(\mathbb{D},\xi_{\mathbb{D}},v_{\mathbb{D}})}(h(b))$. This completes the induction.
		\end{enumerate}
	\end{proof}
	\end{lem}
%	\textbf{Behavioural equivalence}:
	\begin{defn}[Behavioural equivalence]
		Let $(B,\alpha_B,\xi_B,v_B)$ and $(C,\beta_C,\xi_C,v_C)$ be any two $V_{\mathcal{L}}^{bi}$-coalgebra-models, with $(B,\alpha_B)$, $(C,\beta_C)$ objects in $PBS_{\mathcal{L}}$. States $b\in B$ and $c\in C$ are behaviourally equivalent (write $b\approx c$) if there exists a $V_{\mathcal{L}}^{bi}$-coalgebra-model $(Q,\zeta_Q,v_Q)$ and $V_{\mathcal{L}}^{bi}$-coalgebra-model morphisms $h_B:(B,\xi_B,v_B)\to (Q,\zeta_Q,v_Q)$, $h_C:(C,\xi_C,v_C)\to (Q,\zeta_Q,v_Q)$ such that $h_B(b)=h_C(c)$.
	\end{defn}
	The Lindenbaum algebra construction for Fitting's $\mathcal{L}$-valued modal logic: 
	\begin{defn}\label{LAF}
		Define an equivalence relation $\sim$ on $\mathrm{Form}_{\Box}$ by 
		\[
		\varphi \sim \psi \iff \vdash \varphi \leftrightarrow \psi \quad \text{(provable equivalence in the logic)}
		\]
		where $\vdash$ denotes provability in the formal system and $\varphi\leftrightarrow\psi=(\varphi\rightarrow\psi)\wedge(\psi\rightarrow\varphi)$.
		The Lindenbaum algebra is the quotient 
		\[
		A\; :=\; \mathrm{Form}_{\Box}/\!{\sim}=\{[\varphi]:\varphi\in\mathrm{Form}_{\Box}\}
		\]
		where $[\varphi]=\{\psi\in\mathrm{Form}_{\Box}:\varphi\sim\psi\}$ is the equivalence class of $\varphi$, with operations induced by syntax:
		\[
		[\varphi]\vee[\psi]=[\varphi\vee\psi],
		[\varphi]\wedge[\psi]=[\varphi\wedge\psi],
		[\varphi]\rightarrow[\psi]=[\varphi\rightarrow\psi],
		T_{\ell}([\varphi])=[T_\ell(\varphi)],
		\Box([\varphi])=[\Box\varphi],
		\]
		and the nullary operations
		\[
		0_A=[0], \qquad 1_A=[1].
		\]
	\end{defn}
	From the construction of Lindenbaum algebra it follows that $A$ is a $\mathcal{L}$-$\mathcal{ML}$-algebra.

	%\begin{defn}
		%A $PRBS_{\mathcal{L}}$-model $(B,\alpha_B,\mathcal{R},v)$ is image-finite if each state $b\in B$ has finitely many $\mathcal{R}$-successors:
		%\[
		%\forall b\in B,\; \mathcal{R}[b]=\{b'\in B:b\mathcal{R}b'\} \text{ is finite}
		%\]
	%\end{defn}
	%\begin{note}
		%A coalgebraic model $(B,\xi,v)$ is image-finite iff its underlying $V_{\mathcal{L}}^{bi}$-coalgebra $(B,\xi)$ is image-finite, i.e.,
		%\[
		%\forall b\in B,\; \xi(b)\in V_P(B)\ \text{ is a finite subset of } B.
		%\]
		%By the equivalence $PRBS_{\mathcal{L}}\simeq COALG(V_{\mathcal{L}}^{bi})$ and the identification $\xi(b)=\mathcal{R}[b]$, the notions of image-finite $PRBS_{\mathcal{L}}$-model and  image-finite $V_{\mathcal{L}}^{bi}$-coalgebra-model coincide.
	%\end{note}

\begin{thm}\label{BME}
	On $V_{\mathcal{L}}^{bi}$-coalgebra-models (equivalently, on $PRBS_{\mathcal{L}}$-models), behavioural equivalence coincides with modal equivalence.
\begin{proof}
	Let $A$ be the Lindenbaum algebra of Fitting's $\mathcal{L}$-valued modal logic as constructed in Definition \ref{LAF}. Then $A$ is a $\mathcal{L}$-$\mathcal{ML}$-algebra. By our Bitopological duality $\mathcal{MA}_{\mathcal{L}}\simeq PRBS_{\mathcal{L}}^{op}$ (cf. Section \ref{BDFML}), consider its dual space 
	\[
	\mathfrak{X}=\mathcal{G}(A)=(HOM_{\mathcal{VA}_{\mathcal{L}}}(A,\mathcal{L}),\tau_1,\tau_2,\alpha_A,\mathcal{R}_{\Box})
	\]
	where the topology $\tau_1$ is generated by $\langle a\rangle=\{h\in HOM_{\mathcal{VA}_{\mathcal{L}}}(A,\mathcal{L}): h(a)=1\}, a\in A$ and the topology $\tau_2$ whose basis elements are complements of the basis elements of $\tau_1$. The binary relation $\mathcal{R}_{\Box}$ is defined on $HOM_{\mathcal{VA}_{\mathcal{L}}}(A,\mathcal{L})$ as: for any $x,y\in HOM_{\mathcal{VA}_{\mathcal{L}}}(A,\mathcal{L})$, $x\mathcal{R}_{\Box}y\iff \forall a\in A,\ell\in\mathcal{L}\;(x(\Box a)\geq\ell\implies y(a)\geq \ell)$ equivalently, $x\mathcal{R}_{\Box}y\iff \forall a\in A\; (x\in\langle\Box a\rangle \implies y\in\langle a \rangle)$.\\
	Claim$\colon$ for every $a\in A$, $[\mathcal{R}_\Box]\langle a \rangle=\langle\Box a\rangle$. \\
	Let $x\in [\mathcal{R}_\Box]\langle a \rangle$. As $[\mathcal{R}_{\Box}]\langle a \rangle=\{y\in HOM_{\mathcal{VA}_{\mathcal{L}}}(A,\mathcal{L}):\mathcal{R}_{\Box}[y]\subseteq \langle a\rangle\}$, we have $\mathcal{R}_{\Box}[x]\subseteq\langle a\rangle$. Then for every $y\in HOM_{\mathcal{VA}_{\mathcal{L}}}(A,\mathcal{L})$ with $x\mathcal{R}_{\Box}y$ we get $y\in\langle a\rangle$, i.e., $y(a)=1$. Now 
	\[
	x(\Box a)=\bigwedge\{y(a):y\in\mathcal{R}_{\Box}[x]\}
	\]
As $y(a)=1, \forall y\in\mathcal{R}_{\Box}[x]$ and if $\mathcal{R}_{\Box}[x]=\emptyset$, then $\bigwedge\emptyset=1$, hence we have $x(\Box a)=1$.\\
Consider $x\in \langle\Box a\rangle$. Then 
\[
1=x(\Box a)=\bigwedge\{y(a):y\in\mathcal{R}_{\Box}[x]\}
	\]
	So for every $y\in HOM_{\mathcal{VA}_{\mathcal{L}}}(A,\mathcal{L})$ such that $y\in \mathcal{R}_{\Box}[x]$, we have $y(a)=1$. Therefore, $\mathcal{R}_{\Box}[x]\subseteq \langle a\rangle$. Thus, $x\in [\mathcal{R}_\Box]\langle a \rangle$.\\
	On the canonical dual space $\mathfrak{X}=\mathcal{G}(A)$, it is shown that $\tau_1=\tau_2$. By the equivalence $PRBS_{\mathcal{L}}\simeq COALG(V_{\mathcal{L}}^{bi})$ (cf. Theorem \ref{PCI}), we obtain a canonical $V_{\mathcal{L}}^{bi}$-coalgebra structure $(\mathfrak{X},\zeta)$ where
	\[
	\zeta:\mathfrak{X}\to V_{\mathcal{L}}^{bi}(\mathfrak{X})
	\]
	is an arrow in $PBS_{\mathcal{L}}$ with $\zeta(x)=\mathcal{R}_{\Box}[x] \; \text{is a pairwise closed subset of $\mathfrak{X}$}$. For any pairwise closed subset $C$ of $\mathfrak{X}$,
	\[
	\begin{aligned}
	[\zeta]C &= \{x\in HOM_{\mathcal{VA}_{\mathcal{L}}}(A,\mathcal{L}):\zeta(x)\subseteq C\}\\
	              &= \{x\in HOM_{\mathcal{VA}_{\mathcal{L}}}(A,\mathcal{L}):\mathcal{R}_{\Box}[x]\subseteq C\}\\
	              & = [\mathcal{R}_\Box]C.
	            \end{aligned}
	\]
	Therefore, taking $C=\langle a\rangle$, we have 
	\[
	\begin{aligned}
	[\zeta](\langle a\rangle)&=[\mathcal{R}_\Box]\langle a\rangle\\
	&=\langle\Box a\rangle
	\end{aligned}
	\]
	We define a canonical $\mathcal{L}$-valuation $v_\mathfrak{X}:\mathsf{Prop}\to\mathcal{L}^\mathfrak{X}$ on $\mathfrak{X}$ by
	\[
v_\mathfrak{X}(p)(h)=h([p])\qquad (p\in \mathsf{Prop}, h\in HOM_{\mathcal{VA}_{\mathcal{L}}}(A,\mathcal{L}) )
	\]
	For each $p\in\mathsf{Prop}$ and $\ell\in\mathcal{L}$
	\[
	\{h\in HOM_{\mathcal{VA}_{\mathcal{L}}}(A,\mathcal{L}):v_\mathfrak{X}(p)(h)\geq\ell \}=\langle U_\ell([p])\rangle
	\]
	is pairwise clopen in $\mathfrak{X}$ and for every sublagebra $K$ of $\mathcal{L}$, $v_\mathfrak{X}(p)[\alpha_A(K)]\subseteq \alpha_{\mathcal{L}}(K)=K$. Thus $(\mathfrak{X},\zeta,v_\mathfrak{X})$ is a canonical coalgebraic model.\\
	We shall now show that, for every formula $\varphi\in\mathrm{Form}_{\Box}$, and every $h\in HOM_{\mathcal{VA}_{\mathcal{L}}}(A,\mathcal{L})$
	\[
	\llbracket\varphi\rrbracket_{(\mathfrak{X},\zeta,v_\mathfrak{X})}(h)=h([\varphi]).
	\]
	We prove it by induction on $\varphi\in \mathrm{Form}_\Box$.
	\begin{itemize}
 \item For $\varphi=p\in\mathsf{Prop}$, we have $\llbracket p\rrbracket_{(\mathfrak{X},\zeta,v_\mathfrak{X})}(h)=v_X(p)(h)=h([p])$, and $\llbracket 0\rrbracket_{(\mathfrak{X},\zeta,v_\mathfrak{X})}(h)=0=h([0])$, $\llbracket 1\rrbracket_{(\mathfrak{X},\zeta,v_\mathfrak{X})}(h)=1=h([1])$.
 \item Suppose $\llbracket\varphi\rrbracket_{(\mathfrak{X},\zeta,v_\mathfrak{X})}(h)=h([\varphi])$ and 	$\llbracket\psi\rrbracket_{(\mathfrak{X},\zeta,v_\mathfrak{X})}(h)=h([\psi])$ hold for $\varphi$ and $\psi$. Then we have
 \[
 \begin{aligned}
 \llbracket\varphi\star \psi\rrbracket_{(\mathfrak{X},\zeta,v_\mathfrak{X})}(h)&=\llbracket\varphi\rrbracket_{(\mathfrak{X},\zeta,v_\mathfrak{X})}(h)\star \llbracket\psi\rrbracket_{(\mathfrak{X},\zeta,v_\mathfrak{X})}(h)\\
 &=h([\varphi])\star h([\psi]) \; \text{ (by induction hypothesis)}\\
 &=h([\varphi]\star [\psi])\\
 &=h([\varphi\star\psi])\quad (\star\in\{\vee,\wedge,\rightarrow\}),
\end{aligned}
 \]
 and 
 \[
 \begin{aligned}
 \llbracket T_\ell(\varphi)\rrbracket_{(\mathfrak{X},\zeta,v_\mathfrak{X})}&=T_\ell(\llbracket \varphi\rrbracket_{(\mathfrak{X},\zeta,v_\mathfrak{X})}(h))\\
 &=T_\ell(h([\varphi]))\quad \text{ (by induction hypotheisis) }\\
 &=h(T_\ell([\varphi]))\\
 &=h([T_\ell(\varphi)]).
\end{aligned}
 \]
 For the canonical coalgebraic model $(\mathfrak{X},\zeta,v_\mathfrak{X})$, 
 \[
 \llbracket\Box\psi\rrbracket_{(\mathfrak{X},\zeta,v_\mathfrak{X})}(h)=\displaystyle\bigwedge_{y\in\zeta(h)}\llbracket\psi\rrbracket_{(\mathfrak{X},\zeta,v_\mathfrak{X})}(y).
 \]
Let $\ell\in\mathcal{L}$.
 \[
 \begin{aligned}
 \llbracket\Box\psi\rrbracket_{(\mathfrak{X},\zeta,v_\mathfrak{X})}(h)\geq \ell &\iff \forall x\in\zeta(h)\;; \llbracket\psi\rrbracket_{(\mathfrak{X},\zeta,v_\mathfrak{X})}(x)\geq \ell\\
 &\iff \forall x\in\zeta(h)\;; x([\psi])\geq \ell\quad \text{ (by induction hypothesis)}\\
 &\iff \zeta(h)\subseteq\{x\in X:x([\psi])\geq \ell\}=\langle U_a([\psi])\rangle\\
 &\iff h\in [\zeta](\langle U_\ell([\psi])\rangle)\\
 &\iff h\in \langle \Box U_\ell([\psi])\rangle\\
 &\iff h([\Box U_\ell(\psi)])=1\\
 &\iff h([U_\ell(\Box\psi)])=1 \text{ (by the property of modal operator $\Box$)}\\
 &\iff h(U_\ell([\Box\psi]))=1\\
 &\iff h([\Box\psi])\geq \ell.
\end{aligned}
 \]
 Since this holds for every $\ell\in\mathcal{L}$, thus we have 
 \[
 h([\Box\psi])=\llbracket\Box\psi\rrbracket_{(\mathfrak{X},\zeta,v_\mathfrak{X})}(h).
 \]
\end{itemize}
 For any $V_{\mathcal{L}}^{bi}$-coalgebra model $(B,\xi,v)$, we define a theory map $th_B:(B,\xi)\to (\mathfrak{X},\zeta)$ by
 \[
 th_B(b)([\varphi])=\llbracket\varphi\rrbracket_{(B,\xi,v)}(b)\quad (b\in B).\hspace{\fill}{(\ast)}
 \]
 For each $b\in B$, $th_B(b)$ is a $\mathcal{L}$-$\mathcal{VL}$-algebra homomorphism. So, $th_B$ is well-defined. We can now show the following set equality
 \begin{center}
 $\{y\in B:\llbracket\psi\rrbracket_{(B,\xi,v)}(y)=1\}=th_B^{-1}(\{h\in HOM_{\mathcal{VA}_{\mathcal{L}}}(A,\mathcal{L}):h([\psi])=1\})=th_B^{-1}(\langle [\psi]\rangle)$.\hspace{\fill}{$(\ast\ast)$}
\end{center}
 Let $b\in B$. Then 
 \[
 \begin{aligned}
 b\in th_B^{-1}(\langle [\psi]\rangle)&\iff th_B(b)\in \langle [\psi]\rangle\\
 &\iff th_B(b)([\psi])=1\\
 &\iff \llbracket \psi\rrbracket_{(B,\xi,v)}(b)=1\\
 &\iff b\in \{y\in B:\llbracket\psi\rrbracket_{(B,\xi,v)}(y)=1\}
 \end{aligned}
  \]
  We show that $th_B$ is pairwise continuous. It is known that on the canonical space $\mathfrak{X}$ both the topologies equal. Hence, to prove $th_B$ is pairwise continuous, it is sufficient to show that for every basic closed set $\langle a\rangle$ (each $a\in A$, the set $\langle a\rangle$ is clopen in both the topologies on $\mathfrak{X}$) ,
  \[
  th_B^{-1}(\langle a\rangle)\text{ is closed in both the topologies on $B$ }.
  \]
 Let $a\in A$. Choose a formula $\psi\in\mathrm{Form}_\Box$ with $a=[\psi]$. Then,
 \[
 \begin{aligned}
 th_B^{-1}(\langle a\rangle)&=\{b\in B:th_B(b)(a)=1\}\\
 &=\{b\in B:\llbracket\psi\rrbracket_{(B,\xi,v)}(b)=1\}.
\end{aligned}
 \]
 By the induction on formulas and by our model definition, we can show that for every formula $\psi\in\mathrm{Form}_\Box$ and $\ell\in\mathcal{L}$, the set $\{b\in B:\llbracket\psi\rrbracket_{(B,\xi,v)}(b)\geq\ell\}$ is pairwise closed in $B$. Thus, $th_B^{-1}(\langle a\rangle)$ is pairwise closed in $B$, and hence $th_B$ is pairwise continuous.\\
 We show that the theory map $th_B$ is compatible w.r.t. the structure map $\alpha$, i.e., 
 \[
 \text{ for every subalgebra $\mathcal{L}'$ of $\mathcal{L}$},\; th_B(\alpha_B(\mathcal{L}'))\subseteq \alpha_A(\mathcal{L}')=HOM_{\mathcal{VA}_{\mathcal{L}}}(A,\mathcal{L}')
 \]
 We prove it by induction on formulas.
 \begin{enumerate}[(i)]
 	\item Let $p\in\mathsf{Prop}$. If $x\in \alpha_B(\mathcal{L}')$, then $th_B(x)([p])=\llbracket p\rrbracket_{(B,\xi,v)}(x)=v(p)(x)\in\mathcal{L}'$. $\llbracket 0 \rrbracket_{(B,\xi,v)}(x)=0\in\mathcal{L}'$ and $\llbracket 1\rrbracket_{(B,\xi,v)}=1\in\mathcal{L}'$.
 	\item Suppose for $x\in\alpha_B(\mathcal{L}')$, $th_B(x)([\varphi])=\llbracket\varphi\rrbracket_{(B,\xi,v)}(x)\in\mathcal{L}'$ and $th_B(x)([\psi])=\llbracket\psi\rrbracket_{(B,\xi,v)}(x)\in \mathcal{L}'$. Then $\llbracket \varphi\star\psi\rrbracket_{(B,\xi,v)}(x)\in\mathcal{L}'$, where $\star\in\{\vee,\wedge,\rightarrow\}$ and $\llbracket T_\ell(\varphi)\rrbracket_{(B,\xi,v)}(x)\mathcal{L}'$.\\
 	If $x\in\alpha_B(\mathcal{L}')$, then by our model definition we have $\xi(x)\subseteq\alpha_B(\mathcal{L}')$. By induction assumption, for each $y\in\xi(x)$, $th_B(y)([\psi])=\llbracket\psi\rrbracket_{(B,\xi,v)}(y)\in\mathcal{L}'$. Thus
 	\[
 	\llbracket\Box\psi\rrbracket_{(B,\xi,v)}(x)=\displaystyle\bigwedge_{y\in\xi(x)}\llbracket\psi\rrbracket_{(B,\xi,v)}(y)\in\mathcal{L}'\;\text{(Since $\mathcal{L}'$ is the finite Heyting algebra)}
 	\]
 	Therefore, for all $\varphi\in\mathrm{Form}_\Box$, $th_B(x)([\varphi])=\llbracket\varphi\rrbracket_{(B,\xi,v)}(x)\in \mathcal{L}'$, i.e., $th_B(x)\in\alpha_A(\mathcal{L}')$. Hence $th_B(\alpha_B(\mathcal{L}'))\subseteq \alpha_A(\mathcal{L}')$.
 \end{enumerate}
 So, $th_B$ is an arrow in $PBS_{\mathcal{L}}$.\\
 We shall show that $th_B$ is a $V_{\mathcal{L}}^{bi}$-coalgebra morphism and it preserves atomic valuations. In order to show that $th_B$ is a $V_{\mathcal{L}}^{bi}$-coalgebra morphism, we need to verify that 
 \[
 V_{\mathcal{L}}^{bi}(th_B)\circ\xi=\zeta\circ th_B.
 \]
 To prove $th_B$ is a $V_{\mathcal{L}}^{bi}$-coalgebra morphism, it is enough to check that for every pairwise-closed subset $U$ of $\mathfrak{X}$,
 \[
 th_B(b)\in[\zeta]U\iff b\in [\xi](th_B^{-1}(U)).\quad (\ast\ast\ast)
 \]
 We check $(\ast\ast)$ on basic pairwise clopen sets.
 \[
 \begin{aligned}
 	th_B(b)\in [\zeta]\langle a\rangle&\iff \zeta(th_B(b))\subseteq \langle a\rangle\\
 	&\iff th_B(b)\in [\zeta]\langle a\rangle\\
 	&\iff th_B(b)\in\langle\Box a\rangle \;\text{ (as $[\zeta]\langle a\rangle=\langle\Box a\rangle)$}\\
 	&\iff th_B(b)\in \langle\Box [\psi]\rangle\text{ (as $a\in A$, consider $a=[\psi]$, $\psi\in\mathrm{Form}_\Box$) }\\
 	&\iff th_B(b)\in \{h\in HOM_{\mathcal{VA}_{\mathcal{L}}}(A,\mathcal{L}):\llbracket \Box\psi\rrbracket_{(\mathfrak{X},\zeta,v_\mathfrak{X})}(h)=1\}\\
 	&\iff \llbracket\Box\psi\rrbracket_{(\mathfrak{X},\zeta,v_\mathfrak{X})}(th_B(b))=1\\
 	&\iff \llbracket\Box\psi\rrbracket_{(B,\xi,v)}(b)=1 \text{ (using $(\ast)$ and definition of $th_B$)}\\
 	&\iff \displaystyle \bigwedge_{y\in\xi(b)}\llbracket\psi\rrbracket_{(B,\xi,v)}(y)=1\\
 	&\iff \xi(b)\subseteq\{y\in B:\llbracket\psi\rrbracket_{(B,\xi,v)}(y)=1\}\\
 	&\iff b\in [\xi](\{y\in B:\llbracket\psi\rrbracket_{(B,\xi,v)}(y)=1\})\\
 	&\iff b\in [\xi](th_B^{-1}(\{h\in HOM_{\mathcal{VA}_{\mathcal{L}}}(A,\mathcal{L}):h([\psi])=1\}))\text{ (by $(\ast\ast)$)}\\
 	&\iff b\in [\xi](th_B^{-1}(\langle[\psi]\rangle))=[\xi](th_B^{-1}(\langle a \rangle)).
 \end{aligned}
 \]
 For any propositional variable $p\in\mathsf{Prop}$ and $b\in B$, $(v_\mathfrak{X}(p)\circ th_B)(b)=v_\mathfrak{X}(p)(th_B(b))=th_B(b)([p])=\llbracket p\rrbracket_{(B,\xi,v)}(b)=v(p)(b)$. As a result, we have $v_\mathfrak{X}(p)\circ th_B=v(p)$, i.e., $th_B$ preserves atomic valuations. Consequently, $th_B:(B,\xi,v)\to (\mathfrak{X},\zeta,v_{\mathfrak{X}})$ is a $V_\mathcal{L}^{bi}$-coalgebra-model morphism.\\
 Now, we show that modal equivalence concept equivalent to behavioural equivalence. 
 \begin{enumerate}[(a)]
 	\item \textbf{Modal equivalence $\implies$ behavioural equivalence}: Let $(B,\xi,v)$ and $(B',\xi',v')$ be $V_{\mathcal{L}}^{bi}$-coalgebra models. Suppose the states $b\in B$ and $b'\in B'$ are modally equivalent:
 	\[
 	\forall \varphi\in\mathrm{Form}_{\Box}\; :\llbracket\varphi\rrbracket_{(B,\xi,v)}(b)=\llbracket\varphi\rrbracket_{(B',\xi',v')}(b').
 	\]
  The theory maps $th_B:(B,\xi)\to (\mathfrak{X},\zeta)$, $th_{B'}:(B',\xi')\to (\mathfrak{X},\zeta)$ are $V_{\mathcal{L}}^{bi}$-coalgebra morphisms. Then for all $\varphi$,
  \[
  th_B(b)([\varphi])=\llbracket\varphi\rrbracket_{(B,\xi,v)}(b)=\llbracket\varphi\rrbracket_{(B',\xi',v')}(b')=th_{B'}(b')([\varphi]).
  \]
  Since this equality holds for every member of $A$, it follows that $th_B(b)=th_{B'}(b')$. As a result, $b$ and $b'$ are behaviourally equivalent.
 \item \textbf{Behavioural equivalence $\implies$ Modal equivalence}: Let $(B,\xi,v)$ and $(B',\xi',v')$ be $V_{\mathcal{L}}^{bi}$-coalgebra-models. Suppose the states $b\in B$ and $b'\in B'$ are behaviourally equivalent. Then there exists a $V_{\mathcal{L}}^{bi}$-coalgebra-model $(B'',\xi'',v'')$ and  $V_{\mathcal{L}}^{bi}$-coalgebra-model morphisms $f:(B,\xi,v)\to (B'',\xi'',v'')$ and $g:(B',\xi',v')\to (B'',\xi'',v'')$ such that $f(b)=g(b')$. Let $\varphi\in\mathrm{Form}_\Box$. Then, by applying Lemma \ref{VPT} to $V_{\mathcal{L}}^{bi}$-coalgebra-model morphisms $f$ and $g$, we have
 \[
 \llbracket\varphi\rrbracket_{(B,\xi,v)}(b)=\llbracket\varphi\rrbracket_{(B'',\xi'',v'')}(f(b)),
 \]
 and 
 \[
  \llbracket\varphi\rrbracket_{(B',\xi',v')}(b')=\llbracket\varphi\rrbracket_{(B'',\xi'',v'')}(g(b')).
 \]
 Since $f(b)=g(b')$, hence $\llbracket\varphi\rrbracket_{(B,\xi,v)}(b)= \llbracket\varphi\rrbracket_{(B',\xi',v')}(b')$. Since the formula $\varphi$ was chosen arbitrarily, $b$ and $b'$ are modally equivalent. 
 \end{enumerate}
\end{proof} 
\end{thm}
We review the conventional definition of coalgebraic bisimulation concept.
\begin{defn}[\cite{aczel1989}]
	Let $(C_1,T_1)$ and $(C_2,T_2)$ be two $\mathfrak{T}$-coalgebras. A relation $\mathcal{R}\subseteq C_1\times C_2$ is called an Aczel-Mendler bisimulation if there exists a $\mathfrak{T}$-coalgebra $( \mathcal{R},\gamma)$ such that the following diagram 
	\begin{figure}[H]
		\centering
		\begin{tikzcd}
				C_1  \arrow[d, "T_1"'] & \mathcal{R} \arrow[l, "\pi_1"'] \arrow[r, "\pi_2"] \arrow[d, "\gamma"]                       & C_2 \arrow[d, "T_2"] \\
			\mathfrak{T}(C_1)      & \mathfrak{T}(\mathcal{R}) \arrow[l, "\mathfrak{T}(\pi_1)"] \arrow[r, "\mathfrak{T}(\pi_2)"'] & \mathfrak{T}(C_2)   
			\end{tikzcd} 
		\end{figure} 
commutes.
\end{defn}
Now define Aczel-Mendler bisimulations between $V_{\mathcal{L}}^{bi}$-coalgebra-models.
\begin{defn}[$V_{\mathcal{L}}^{bi}$-bisimulation]\label{VCB}
	Let $(B,\alpha_B,\xi_B,v_B)$ and $(C,\beta_C,\xi_C,v_C)$ be any two $V_{\mathcal{L}}^{bi}$-coalgebra-models, with $(B,\alpha_B)$ and $(C,\beta_C)$ objects in $PBS_{\mathcal{L}}$. A relation $\mathcal{Z}\subseteq B\times C$ is said to be $V_{\mathcal{L}}^{bi}$-bisimulation if for all $(b,c)\in\mathcal{Z}$ and each $p\mathsf{Prop}$, we have $v_B(p)(b)=v_C(p)(c)$, and if there exists a coalgebraic structure $\zeta:(\mathcal{Z},\alpha_{\mathcal{Z}})\to V_{\mathcal{L}}^{bi}(\mathcal{Z},\alpha_{\mathcal{Z}})$, where $\mathcal{Z}$ is a pairwise closed subspace of $B\times C$ and $\alpha_{\mathcal{Z}}:\mathfrak{S}_{\mathcal{L}}\to\Lambda_{\mathcal{Z}}$ is a subalgebra-indexed, meet-preserving structure map defined by $\alpha_{\mathcal{Z}}(A)=\mathcal{Z}\cap (\alpha_B(A)\times \alpha_C(A))$, hence $(\mathcal{Z},\alpha_{\mathcal{Z}})$ is an object in $PBS_{\mathcal{L}}$, such that the projection maps $\pi_B:\mathcal{Z}\to B$ and $\pi_C:\mathcal{Z}\to C$, which are morphisms in $PBS_{\mathcal{L}}$ that is pairwise continuous and structure map-compatible, i.e., $\pi_B^{-1}(\alpha_B(A))=\alpha_{\mathcal{Z}}(A)=\pi_C^{-1}(\beta_C(A))$, $\forall A\in \mathfrak{S}_{\mathcal{L}}$, are $V_{\mathcal{L}}^{bi}$-coalgebra morphisms:
\[
\begin{aligned}
	\xi_B\circ\pi_B&=V_{\mathcal{L}}^{bi}(\pi_B)\circ\zeta,\\ V_{\mathcal{L}}^{bi}(\pi_C)\circ\zeta&=\xi_C\circ\pi_C.
\end{aligned}
\]
\end{defn}
Two states $b\in B$ and $c\in C$ are said to be bisimilar if there exists a bisimulation $\mathcal{Z}$ such that $(b,c)\in \mathcal{Z}$.
We write $b\leftrightarroweq_{\mathcal{Z}} c$ if $(b,c)\in Z$, and $b\leftrightarroweq c$ if the states $b$ and $c$ are bisimilar.
The following definition is equivalent to the coalgebraic bisimulation given in Definition \ref{VCB}; it is stated in the relational language of $PRBS_{\mathcal{L}}$.
\begin{defn}
	A binary relation $\mathcal{Z}\subseteq B\times C$ is a bisimulation between any two objects $(B,\alpha_B,\mathcal{R},v_B)$ and $(C,\beta_C,\mathcal{S},v_C)$ in $PRBS_{\mathcal{L}}$ if 
	\begin{enumerate}
		\item for any $(b,c)\in\mathcal{Z}$ and each $p\in\mathsf{Prop}$, $v_B(p)(b)=v_C(p)(c)$;
		\item $\mathcal{Z}$ is a pairwise closed subspace of $B\times C$, and for all $(b,c)\in\mathcal{Z}$ and $\mathcal{L}'\in \mathfrak{S}_{\mathcal{L}}$, $b\in\alpha_B(\mathcal{L}')\iff c\in \beta_C(\mathcal{L}')$;
		\item for any $(x,y)\in \mathcal{Z}$ and $x'\in \mathcal{R}[x]$, then there exists $y'\in\mathcal{S}[y]$ such that $(x',y')\in \mathcal{Z}$;
		\item for any $(x,y)\in\mathcal{Z}$ and $y'\in\mathcal{S}[y]$, then there exists $x'\in \mathcal{R}[x]$ such that $(x',y')\in\mathcal{Z}$.
	\end{enumerate}
\end{defn}

\begin{thm}[Hennessy-Milner theorem for $V_{\mathcal{L}}^{bi}$-coalgebra-models]
	\label{MEB}
	Let $(B,\xi,v)$ and $(C,\gamma,w)$ be two $V_{\mathcal{L}}^{bi}$-coalgebra-models. For any two states $b\in B$ and $c\in C$
		\[
	b\equiv c\iff b\leftrightarroweq c
		\]
		\begin{proof}
			Suppose $b\leftrightarroweq c$. Then there exists a $V_{\mathcal{L}}^{bi}$-bisimulation $\mathcal{Z}$ between $V_{\mathcal{L}}^{bi}$-coalgebra-models $(B,\xi,v)$ and $(C,\gamma,w)$ such that $(b,c)\in \mathcal{Z}$. We show that the states $b$ and $c$ are modally equivalent. Now $(\mathcal{Z},\zeta_{\mathcal{Z}},v_{\mathcal{Z}})$ is the $V_{\mathcal{L}}^{bi}$-coalgebra-model with valuation $v_{\mathcal{Z}}(p)=v(p)\circ \pi_1=w(p)\circ\pi_2$. Then the projection maps $\pi_1:(\mathcal{Z}\zeta_{\mathcal{Z}},v_{\mathcal{Z}})\to (B,\xi,v)$ and $\pi_2:(\mathcal{Z}\zeta_{\mathcal{Z}},v_{\mathcal{Z}})\to (C,\gamma,w)$ are $V_{\mathcal{L}}^{bi}$-colagebra-model morphisms. By Lemma \ref{VPT}, we have 
				\[
				\llbracket\varphi\rrbracket_{(\mathcal{Z},\zeta_{\mathcal{Z}},v_{\mathcal{Z}})}=\llbracket\varphi\rrbracket_{(B,\xi,v)}(\pi_1(z)),
				\]
				and 
				\[
				\llbracket\varphi\rrbracket_{(\mathcal{Z},\zeta_{\mathcal{Z}},v_{\mathcal{Z}})}=\llbracket\varphi\rrbracket_{(C,\gamma,w)}(\pi_2(z)),
				\]
				for $z\in\mathcal{Z}$ with $\pi_1(z)=b$ and $\pi_2(z)=c$.
				Thus, we have 
				\[
				\llbracket\varphi\rrbracket_{(B,\xi,v)}(b)=\llbracket\varphi\rrbracket_{(C,\gamma,w)}(c).
				\]
				Hence the states $b$ and $c$ are modally equivalent.\\
				
				Let $b\equiv c$. We show that $b\leftrightarroweq c$. The approach employed in the proof of Theorem \ref{BME} shows that the theory maps $th_B:(B,\xi)\to (\mathfrak{X},\zeta)$, $th_C:(C,\gamma)\to (\mathfrak{X},\zeta)$ are $V_{\mathcal{L}}^{bi}$-coalgebra morphisms. As $b\equiv c$,
				\[
				\forall \varphi\in\mathrm{Form}_\Box\colon th_B(b)([\varphi])=\llbracket\varphi\rrbracket_{(B,\xi,v)}(b)=\llbracket\varphi\rrbracket_{(C,\gamma,w)}(c)=th_C(c)([\varphi]).
				\]
				As a result, $th_B(b)=th_C(c)$. Define the relation $\mathcal{Z}=\{(x,y)\in B\times C:th_B(x)=th_C(y)\}$. Then $(b,c)\in\mathcal{Z}$. The map $th_B\times th_C:B\times C\to \mathfrak{X}\times\mathfrak{X}$ is an arrow in $PBS_{\mathcal{L}}$, and the diagonal $\nabla=\{(h,h):h\in HOM_{\mathcal{VA}_{\mathcal{L}}}(A,\mathcal{L}) \}$ is pairwise closed in $\mathfrak{X}\times\mathfrak{X}$. So, $(th_B\times th_C)^{-1}(\nabla)=\mathcal{Z}$ is pairwise closed in $B\times C$. Define $\alpha_{\mathcal{Z}}(\mathcal{L}')=\mathcal{Z}\cap(\alpha_B(\mathcal{L}')\times\alpha_C(\mathcal{L}'))$, where $\mathcal{L}'\in\mathfrak{S}_{\mathcal{L}}$. Thus, $(\mathcal{Z},\alpha_{\mathcal{Z}})$ is an object in $PBS_{\mathcal{L}}$.\\
				For any $(x,y)\in\mathcal{Z}$ and $p\in\mathsf{Prop}$, 
				\[
				v(p)(x)=th_B(x)([p])=th_C(y)([p])=w(p)(y).
				\]
				Let $\pi_1:\mathcal{Z}\to B$ and $\pi_2:\mathcal{Z}\to C$ be the projection maps. Then both the projection maps $\pi_1$, $\pi_2$ are arrows in $PBS_{\mathcal{L}}$.
				Define $v_{\mathcal{Z}}(p)=v(p)\circ \pi_1=w(p)\circ \pi_2$. As $th_B$, $th_C$ are coalgebra morphisms, we have 
				\[
				V_{\mathcal{L}}^{bi}(th_B)\circ\xi=\zeta\circ th_B,\; V_{\mathcal{L}}^{bi}(th_C)\circ\gamma=\zeta\circ th_C.
				\]
				For any $(x,y)\in\mathcal{Z}$, we have \\
				$V_{\mathcal{L}}^{bi}(th_B)(\xi(x))=\zeta(th_B(x))=\zeta(th_C(y))=V_{\mathcal{L}}^{bi}(th_C)(\gamma(y))$. \\
				By the notion of endofunctor $V_{\mathcal{L}}^{bi}$, the above equality gives:
				\[
				\{th_B(x^*):x^*\in\xi(x)\}=\{th_C(y^*):y^*\in\gamma(y)\}.\qquad \qquad{(\dagger)}
				\]
				Define $\zeta_{\mathcal{Z}}:\mathcal{Z}\to V_{\mathcal{L}}^{bi}(\mathcal{Z})$ by
				\[
				\zeta_{\mathcal{Z}}(x,y)=\mathcal{Z}\cap(\xi(x)\times\gamma(y)).
				\]
				As $\mathcal{Z}$ and $\xi(x)\times\gamma(y)$ are pairwise closed in $B\times C$, for each $(x,y)\in\mathcal{Z}$, $\zeta_{\mathcal{Z}}(x,y)$ is pairwise closed in $B\times C$ and it is pairwise closed in $\mathcal{Z}$ as well. Therefore, $\zeta_{\mathcal{Z}}(x,y)\in V_{\mathcal{L}}^{bi}(\mathcal{Z})$. Now we show that $\zeta_{\mathcal{Z}}$ is an arrow in $PBS_{\mathcal{L}}$. We prove this by factoring $\zeta_{\mathcal{Z}}$ as a composition of arrows in $PBS_{\mathcal{L}}$.
				\begin{enumerate}[(i)]
			\item 	As both $\xi$ and $\gamma$ are arrows in $PBS_{\mathcal{L}}$, so
				\[
				\xi\times\gamma:B\times C\to V_{\mathcal{L}}^{bi}(B)\times V_{\mathcal{L}}^{bi}(C)
				\]
				is an arrow in $PBS_{\mathcal{L}}$.
				\item The inclusion map $\mathfrak{i}:\mathcal{Z}\to B\times C$ is an arrow in $PBS_{\mathcal{L}}$. 
				\item Define a mapping $\mathfrak{m}:V_{\mathcal{L}}^{bi}(B)\times V_{\mathcal{L}}^{bi}(C)\to V_{\mathcal{L}}^{bi}(B\times C)$ by $\mathfrak{m}(U,V)=U\times V$. It is easy to follow that $\mathfrak{m}$ is well-defined. If $M\in\beta_1^B$ and $N\in\beta_1^C$, then 
				\[
				\begin{aligned}
				\mathfrak{m}^{-1}(\Box (M\times N))&=\{(B',C')\in V_P(B)\times V_P(C):B'\times C'\subseteq M\times N \}\\
				&=\{(B',C')\in V_P(B)\times V_P(C):B'\subseteq M \& C'\subseteq N\}\\
				&=\Box M\times \Box N,
				\end{aligned}
				\]
				and 
				\[
				\begin{aligned}
			\mathfrak{m}^{-1}(\Diamond (M\times N))&=\{(B',C')\in V_P(B)\times V_P(C):(B'\times C')\cap (M\times N)\neq\emptyset \}\\
			&=\{(B',C')\in V_P(B)\times V_P(C):B'\cap M\neq\emptyset \& C'\cap N\neq\emptyset\}\\
			&=\Diamond M\times \Diamond N.
		\end{aligned}
				\]
				Similarly, if $M\in\beta_2^B$ and $N\in\beta_2^C$, then $\mathfrak{m}^{-1}(\Box (M\times N))=\Box M\times \Box N$ and $\mathfrak{m}^{-1}(\Diamond (M\times N))=\Diamond M\times \Diamond N$. Thus $\mathfrak{m}$ is a pairwise continuous map. $\mathfrak{m}$ is also structure map compatible because if $U\in V_P\circ \alpha_B(\mathcal{L}')$ and $V\in V_P\circ \alpha_C(\mathcal{L}')$, where $\mathcal{L}'\in\mathfrak{S}_{\mathcal{L}}$, i.e.,  $U\subseteq\alpha_B(\mathcal{L}')$ and $V\subseteq\alpha_C(\mathcal{L}')$, then $U\times V\subseteq \alpha_B(\mathcal{L}')\times\alpha_C(\mathcal{L}')=\alpha_{B\times C}(\mathcal{L}')$ i.e., $\mathfrak{m}(U,V)\in V_P\circ\alpha_{B\times C}(\mathcal{L}')$. Hence, $\mathfrak{m}$ is an arrow in $PBS_{\mathcal{L}}$.
				\item Define a map $j_{\mathcal{Z}}:V_{\mathcal{L}}^{bi}(B\times C)\to V_{\mathcal{L}}^{bi}(\mathcal{Z})$ by $j_{\mathcal{Z}}(\mathfrak{W})=\mathcal{Z}\cap\mathfrak{W}$. As $\mathcal{Z}$ is a pairwise closed subset of $B\times C$, thus for any $\mathfrak{W}\in V_P(B\times C)$, $\mathcal{Z}\cap\mathfrak{W}$ is pairwise closed in the subspace $\mathcal{Z}$, i.e., $j_{\mathcal{Z}}(\mathfrak{W})\in V_P(\mathcal{Z})$. Let $U\in\beta_1^{\mathcal{Z}}$ and $V\in\beta_2^{\mathcal{Z}}$. Then $U=\mathcal{Z}\cap W$, $W\in\beta_1^{B\times C}$.
				\[
				\begin{aligned}
				j_{\mathcal{Z}}^{-1}(\Box U)&=\{W^*\in V_P(B\times C):j_{\mathcal{Z}}(W^*)\in\Box U\}\\
				&=\{W^*\in V_P(B\times C):\mathcal{Z}\cap W^*\subseteq U\}\\
				&=\{W^*\in V_P(B\times C):\mathcal{Z}\cap W^*\subseteq \mathcal{Z}\cap W\}\\
				&=\{W^*\in V_P(B\times C):W^*\subseteq W\cup \mathcal{Z}^c\}\\
				&=\Box(W\cup \mathcal{Z}^c),
			\end{aligned}
				\]
				and 
				\[
				\begin{aligned}
				j_{\mathcal{Z}}^{-1}(\Diamond U)&=\{W^*\in V_P(B\times C):(\mathcal{Z}\cap W^*)\cap (\mathcal{Z}\cap W)\neq\emptyset\}\\
				&=\{ W^*\in \mathcal{K}(B\times C): W^*\cap W\neq\emptyset    \}\\
				&=\Diamond W.
			\end{aligned}
				\]
				Similar equalities hold for the case when $V\in\beta_2^{\mathcal{Z}}$. Thus, $j_{\mathcal{Z}}$ is a pairwise continuous. \\
				Let $\mathcal{L}'\in\mathfrak{S}_{\mathcal{L}}$ and $S\in V_P\circ \alpha_{B\times C}(\mathcal{L}')$. Then $S\subseteq \alpha_{B\times C}(\mathcal{L}')$ and $j_{\mathcal{Z}}(S)=\mathcal{Z}\cap S\subseteq \mathcal{Z}\cap \alpha_{B\times C}(\mathcal{L}')=\alpha_{\mathcal{Z}}(\mathcal{L}')$. Thus, we have $j_{\mathcal{Z}}(V_P\circ \alpha_{B\times C}(\mathcal{L}'))\subseteq V_P\circ \alpha_{\mathcal{Z}}(\mathcal{L}')$. Therefore, $j_{\mathcal{Z}}$ is an arrow in $PBS_{\mathcal{L}}$.
			\end{enumerate}
			Now, $\zeta_{\mathcal{Z}}=j_{\mathcal{Z}}\circ \mathfrak{m}\circ \xi\times\gamma\circ\mathfrak{i}:\mathcal{Z}\to V_{\mathcal{L}}^{bi}(\mathcal{Z})$ is an arrow in $PBS_{\mathcal{L}}$. Henceforth, $(\mathcal{Z},\zeta_{\mathcal{Z}},v_{\mathcal{Z}})$ is a $V_{\mathcal{L}}^{bi}$-coalgebra-model. it is now required to show that the projection maps $\pi_1$ and $\pi_2$ are $V_{\mathcal{L}}^{bi}$-coalgebra morphisms. We show that 
			\[
			V_{\mathcal{L}}^{bi}(\pi_1)\circ\zeta_{\mathcal{Z}}=\xi\circ\pi_1,\; V_{\mathcal{L}}^{bi}(\pi_2)\circ\zeta_{\mathcal{Z}}=\gamma\circ\pi_2.
			\]
			Let $(x,y)\in\mathcal{Z}$. Then 
			\[
			\begin{aligned}
			V_{\mathcal{L}}^{bi}(\pi_1)(\zeta_{\mathcal{Z}}(x,y))&=V_{\mathcal{L}}^{bi}(\pi_1)(\mathcal{Z}\cap(\xi(x)\times\gamma(y)))\\
			&=\pi_1[\mathcal{Z}\cap(\xi(x)\times\gamma(y))]\subseteq\xi(x).
		\end{aligned}
			\]
			From $(\dagger)$, we have $\forall x^*\in\xi(x)\exists y^*\in\gamma(y) \text{ such that } th_B(x^*)=th_C(y^*)$, i.e., $(x^*,y^*)\in\mathcal{Z}$. Hence, $\xi(x)\subseteq \pi_1[\mathcal{Z}\cap(\xi(x)\times\gamma(y))]$. Thus, $V_{\mathcal{L}}^{bi}(\pi_1)\circ\zeta_{\mathcal{Z}}=\xi\circ\pi_1$. Similarly, we have $V_{\mathcal{L}}^{bi}(\pi_2)\circ\zeta_{\mathcal{Z}}=\gamma\circ\pi_2$. Therefore, both the projection maps $\pi_1$ and $\pi_2$ are coalgebra morphisms. Hence, $\mathcal{Z}$ is a $V_{\mathcal{L}}^{bi}$-bisimulation.
			%	Let $U\in\beta_1^{\mathcal{Z}}$. Then $U=\mathcal{Z}\cap W$, where $W\in\beta_1^{(B\times C)}$. Now,
				%\[
				%\begin{aligned}
				%	\zeta_{\mathcal{Z}}^{-1}(\Diamond U)&=\{(x,y)\in\mathcal{Z}:\zeta_{\mathcal{Z}}(x,y)\cap U\neq\emptyset\}\\
					%&=\{(x,y)\in\mathcal{Z}:(\mathcal{Z}\cap(\xi(x)\times\gamma(y)))\cap ()\neq \emptyset\}\\
					%&=\{(x,y)\in\mathcal{Z}:(\xi(x)\times\gamma(y))\cap U\neq\emptyset\}\;\text{ ( as $U\subseteq \mathcal{Z})$}\\
					%&=\{(x,y)\in\mathcal{Z}:(\xi(x)\times\gamma(y))\cap (\mathcal{Z}\cap W)\neq\emptyset\}
				%\end{aligned}
				
				%\]

		\end{proof}
\end{thm}
By combining Theorem \ref{BME} and Theorem \ref{MEB}, we get the following result:
\begin{cor}\label{BHCB}
	On $V_{\mathcal{L}}^{bi}$-coalgebra-models, behavioural equivalence coincides with $V_{\mathcal{L}}^{bi}$-bisimulation.
\end{cor}

We now show the structural consequences of our established dualities.
\begin{cor}
	The category $COALG(V_{\mathcal{L}}^{bi})$ has cofree coalgebra structure.
	\begin{proof}
	%	Let $\mathcal{U}_{COALG(V_{\mathcal{L}}^{bi})}:COALG(V_{\mathcal{L}}^{bi})\to PBS_{\mathcal{L}}$ and $U_{PRBS_{\mathcal{L}}:PRBS_{\mathcal{L}}\to PBS_{\mathcal{L}}$ be the forgetful functors. As  $PRBS_{\mathcal{L}}\cong COALG(V_{\mathcal{L}}^{bi})$, and $\mathcal{B}:PRBS_{\mathcal{L}}\to COALG(V_{\mathcal{L}}^{bi})$ is the category isomorphism (cf. Theorem \ref{}), hence $\mathcal{U}_{COALG(V_{\mathcal{L}}^{bi})}=U_{PRBS_{\mathcal{L}}\circ \mathcal{B}^{-1}$. So
			%	\[
				% \mathcal{U}_{COALG(V_{\mathcal{L}}^{bi})}\; \text{has a right adjoint}\iff U_{PRBS_{\mathcal{L}}\;\text{ has a right adjoint}.
		%	\]
		%	Define a functor $\mathfrak{E}:PBS_{\mathcal{L}}\to PRBS_{\mathcal{L}}$ by
		%	\begin{enumerate}[(i)]
				%\item for each object $(B,\alpha_B)$ in $PBS_{\mathcal{L}}$, $\mathfrak{E}((B,\alpha_B))=(B,\alpha_B^*,\mathcal{R}^{\emptyset})$, where $\alpha_B^*(\mathcal{L}')=\mathcal{L}'$,$\forall \mathcal{L}'\in\mathfrak{S}_{\mathcal{L}}$;
			%	\item for an arrow $f: (B,\alpha_B)\to (C,\alpha_C)$ in $PBS_{\mathcal{L}}$, define $\mathcal{R}(f):\mathcal{R}(B)\to\mathcal{R}(C)$ by $\mathcal{R}(f)=f$.
		%	\end{enumerate}
			
			Let $\mathcal{U}_{COALG(V_{\mathcal{L}}^{bi})}:COALG(V_{\mathcal{L}}^{bi})\to PBS_{\mathcal{L}}$ be the forgetful functor. To prove the existence of cofree $V_{\mathcal{L}}^{bi}$-coalgebra, we need to show that the forgetful functor $\mathcal{U}_{COALG(V_{\mathcal{L}}^{bi})}$ has a right adjoint. We define the forgetful functor $U_\Box:\mathcal{MA}_{\mathcal{L}}\to\mathcal{VA}_{\mathcal{L}}$ as follows:
			\begin{enumerate}[(i)]
				\item On objects: for an object $(A,\Box)$ in $\mathcal{MA}_{\mathcal{L}}$, where $A$ is a $\mathcal{L}$-$\mathcal{VL}$-algebra, $U_\Box((A,\Box))=A$;
				\item On arrows: for an arrow $h$ in $\mathcal{MA}_{\mathcal{L}}$, $U_\Box(h)=h \;\text{(underlying $\mathcal{L}$-$\mathcal{VL}$-algebra homomorphism)}$.
			\end{enumerate}
		Since $\mathcal{L}$-$\mathcal{ML}$-algebra is a $\mathcal{L}$-$\mathcal{VL}$-algebra together with a modal operator $\Box$ satisfying the axioms mentioned in Definition \ref{LML}, so the functor $U_\Box$ is well-defined. \\
		The category $\mathcal{VA}_{\mathcal{L}}$ of $\mathcal{L}$-$\mathcal{VL}$-algebras is a variety. So free algebra exists in $\mathcal{VA}_{\mathcal{L}}$. We now define the free functor $F_\Box:\mathcal{VA}_{\mathcal{L}}\to \mathcal{MA}_{\mathcal{L}}$ as follows:
		\begin{enumerate}[(i)]
			\item On objects: for an object $A$ in $\mathcal{VA}_{\mathcal{L}}$, $F_\Box(A)$ is the $\mathcal{L}$-$\mathcal{ML}$-algebra freely generated by $A$ under the modal operation $\Box$ subject to 
			\[
			\Box(x\wedge y)=\Box x\wedge\Box y, \Box U_\ell(x)=U_\ell(\Box x),
			\]
			where $x,y\in F_\Box(A)$, i.e., $F_\Box(A)$ is the quotient of the term algebra in the signature $\{\vee,\wedge,T_\ell(\ell\in\mathcal{L}),\rightarrow,\Box,0,1\}$ with all nullary function symbols $c_a, a\in A$, by the smallest congruence generated by:
			\begin{enumerate}[(i)]
				\item all the $\mathcal{L}$-$\mathcal{VL}$-algebra axioms;
				\item the modal $\Box$-axioms: for all terms $x,y$, $\Box(x\wedge y)=\Box x\wedge\Box y$, $U_\ell(\Box x)=\Box(U_\ell(x))$, $\ell\in\mathcal{L}$;
				\item the relations: $\mathfrak{g}(c_{a_1},c_{a_2},\cdots,c_{a_n})=c_{\mathfrak{g}(a_1,a_2,\cdots,a_n)}$ for every operation $\mathfrak{g}\in\{\vee,\wedge,T_\ell(\ell\in\mathcal{L}),\rightarrow,0,1\}$ of arity $n$.
				\end{enumerate}

			\item On arrows: for an arrow $f:A\to B$ in $\mathcal{VA}_{\mathcal{L}}$, define $F_\Box(f):F_\Box(A)\to F_\Box(B)$ by $F_\Box(f)([t(c_{a_1},c_{a_2},\cdots,c_{a_n})])=[t(c_{f(a_1)},c_{f(a_2)},\cdots,c_{f(a_n)})]$, where $a\in A$ and $[t(c_{a_1},c_{a_2},\cdots,c_{a_n})]$ denotes the equivalence class of the base term $t(c_{a_1},c_{a_2},\cdots,c_{a_n})$.
		\end{enumerate}
			For each object $A$ in $\mathcal{VA}_{\mathcal{L}}$, there is a canonical $\mathcal{L}$-$\mathcal{VL}$-algebra homomorphism $\eta_A:A\to U_\Box(F_\Box(A))$ defined by $\eta_A(a)=[c_a]$. Then it can be shown that, for each object $M$ in $\mathcal{MA}_{\mathcal{L}}$ and each morphism $h: A\to U_\Box(M)$ in $\mathcal{VA}_{\mathcal{L}}$ there exists a unique morphism $\hat{h}:F_\Box(A)\to M$ in $\mathcal{MA}_{\mathcal{L}}$ such that $U_\Box(\hat{h})\circ\eta_A=h$. So, $\eta_A$ is the unit and $F_\Box$ is left adjoint to $U_\Box$. \\
			Using the dual equivalences $\mathcal{VA}_{\mathcal{L}}\simeq PBS_{\mathcal{L}}^{op}$, $\mathcal{MA}_{\mathcal{L}}\simeq PRBS_{\mathcal{L}}^{op}$ and $PRBS_{\mathcal{L}}\cong COALG(V_{\mathcal{L}}^{bi})$ together with the adjunction $F_\Box\dashv U_\Box$, we define a functor 	$\mathcal{H}=\mathfrak{B}\circ\mathcal{G}\circ F_\Box\circ \mathfrak{F}:PBS_{\mathcal{L}}\to COALG(V_{\mathcal{L}}^{bi})$, where 
			\begin{enumerate}[(i)]
				\item $\mathfrak{B}:PRBS_{\mathcal{L}}\to COALG(V_{\mathcal{L}}^{bi})$ is the covariant functor,
				\item $\mathcal{G}:\mathcal{MA}_{\mathcal{L}}\to PRBS_{\mathcal{L}}$ is the contravariant functor,
				\item $F_\Box:\mathcal{VA}_{\mathcal{L}}\to \mathcal{MA}_{\mathcal{L}}$ is the free functor,
				\item $\mathfrak{F}:PBS_{\mathcal{L}}\to\mathcal{VA}_{\mathcal{L}}$ is the contravariant functor,
			\end{enumerate}
			by:	
			\begin{enumerate}[(i)]
		\item 	for an object $(B,\alpha_B)$ in $PBS_{\mathcal{L}}$, $\mathcal{H}((B,\alpha_B))$ is the $V_{\mathcal{L}}^{bi}$-coalgebra $(P,\zeta)$ obtained by applying $\mathfrak{B}$ to the object $\mathfrak{F}(F_\Box(\mathcal{G}(B)))$ in $PRBS_{\mathcal{L}}$ with $[\zeta](\langle p\rangle)=\langle\Box p\rangle$,
		\item for an arrow $f$ in $PBS_{\mathcal{L}}$, $\mathcal{H}(f)=\mathfrak{F}(F_\Box(\mathcal{G}(f)))$.
	\end{enumerate}
	The counit $\xi_B:\mathcal{U}_{COALG(V_{\mathcal{L}}^{bi})}(\mathcal{H}(B))\to B$ shows that $\mathcal{H}$ is right adjoint to $\mathcal{U}_{COALG(V_{\mathcal{L}}^{bi})}$. Thus, for every object in $PBS_{\mathcal{L}}$, there exists a cofree $V_{\mathcal{L}}^{bi}$-coalgebra structure.

	\end{proof}
	\end{cor}
\begin{cor}\label{FCF}
A final coalgebra exists for the endofunctor $V_{\mathcal{L}}^{bi}$.
\begin{proof}
	The category $\mathcal{MA}_{\mathcal{L}}$ of $\mathcal{L}$-$\mathcal{ML}$-algebras is a variety. Hence, by the help of standard results in universal algebra, $\mathcal{MA}_{\mathcal{L}}$ admits free algebras on all sets of generators. Thus, there exists a free algebra on the set $\mathsf{Prop}=\emptyset$ and it meets the criterion for an initial object in the category $\mathcal{MA}_{\mathcal{L}}$. By the established dual equivalence
	\[
	\mathcal{MA}_{\mathcal{L}}\simeq COALG(V_{\mathcal{L}}^{bi})^{op},
	\]
	initial objects in $\mathcal{MA}_{\mathcal{L}}$ correspond to terminal objects in the category $COALG(V_{\mathcal{L}}^{bi})$. A terminal object in $COALG(V_{\mathcal{L}}^{bi})$ is actually a final coalgebra for the endofunctor $V_{\mathcal{L}}^{bi}$. As $\mathcal{MA}_{\mathcal{L}}$ has an initial algebra, it is clear that final coalgebra exists for $V_{\mathcal{L}}^{bi}$.
\end{proof}
\end{cor}
%\begin{rem}
	%With the help of Aczel-Mendler bisimulation notion, bisimilarity is the kernel of the unique morphism into the final coalgebra
%\end{rem}

\section{Conclusion}
\label{FCON}
We have introduced the category $PRBS_{\mathcal{L}}$ and found a duality for the class of all algebras of a version of Fitting's Heyting-valued modal logic in bitopological language. This has led to an extension of the natural duality theory for modal algebras. We have demonstrated how the theory of coalgebras can be used to characterise the category $PRBS_{\mathcal{L}}$ and thus obtained a coalgebraic description of the bitopological duality for Fitting's Heyting-valued modal logic. We have adapted Lauridsen's bi-Vietoris functor construction \cite{lauridsen2015bitopological} to our $\mathcal{L}$-valued pairwise Boolean spaces setting and used it to obtain the coalgebraic duality for Fitting's many-valued modal logic.
%The Vietoris functor on the category $PBS_{\mathcal{L}}$ has been formally constructed in the present study, and we have ultimately shown that coalgebras for this functor provide sound and complete semantics for Fitting's multi-valued modal logic. 
As an application of this coalgebraic duality, we have established the Hennessy-Milner theorem for Fitting's $\mathcal{L}$-valued modal logic. Furthermore, we have shown the existence of a final coalgebra and cofree coalgebras in the category $COALG(V_{\mathcal{L}}^{bi})$.\\
We work with finite Heyting algebra $\mathcal{L}$ and $\mathcal{L}$-valued pairwise Boolean spaces. We treat one unary modal operation $\Box$, Boolean negation and $\Diamond$ are not considered as primitive in our framework, and we do not impose any conditions (e.g., reflexivity, transitivity) on $\mathcal{L}$-valued Kripke frame. Extensions to multiple modalities, graded or conditional modalities are not addressed in this study. The existence of the final and cofree coalgebra in $COALG(V_{\mathcal{L}}^{bi})$ has been shown via the established dualities and adjunctions. We do not provide constructive descriptions or computational consequences for these coalgebras. Extensions of our established dualities, the Hennessy-Milner theorem, and the coalgebraic constructions to infinite, non-distributive or residuated $\mathcal{L}$ remain open.\\
In light of our current work, we can suggest some future study directions.\\
A promising avenue for future research could involve demonstrating how lattice-valued intuitionistic modal logic can be characterized by the coalgebras of an endofunctor $V$ on the category $BES$ of bi-topological Esakia spaces (the concept of bitopological Esakia spaces is discussed in \cite{bezhanishvili2010bitopological}). We are still determining, though, how to describe the relation $\mathcal{R}$ on bitopological Esakia spaces in terms of coalgebras of the functor $V$; this seems to be a topic of inquiry at present.
\section*{Declaration of Competing Interest}
The authors declare that they have no known competing financial interests or personal relationships that could have appeared to influence the work reported in this paper.
\section*{Acknowledgements}
I express my gratitude to Kumar Sankar Ray and Prakash Chandra Mali for their insightful comments and discussions. This research received no external funding.
\newpage
%\section*{References}
	
\end{document}